\documentclass[letterpaper,aps,prd,preprint,showpacs,nofootinbib,
superscriptaddress]{revtex4}
\pdfoutput=1
\usepackage{epsf}
\usepackage{epsfig}
\usepackage{graphics}
\usepackage{graphicx}
\usepackage{amssymb}

\usepackage{color}

\newcommand{\nc}{\newcommand}
\nc{\postscript}[2]{\setlength{\epsfxsize}{#2\hsize}\centerline{\epsfbox{#1}}}

\nc{\beq}{\begin{equation}}   \nc{\eeq}{\end{equation}}
\nc{\bea}{\begin{eqnarray}}   \nc{\eea}{\end{eqnarray}}
\nc{\baa}{\begin{array}}      \nc{\eaa}{\end{array}}
\nc{\bit}{\begin{itemize}}    \nc{\eit}{\end{itemize}}
\nc{\ben}{\begin{enumerate}}  \nc{\een}{\end{enumerate}}
\nc{\bce}{\begin{center}}     \nc{\ece}{\end{center}}

\def\lsim{\mathrel{\raise.3ex\hbox{$<$\kern-.75em\lower1ex\hbox{$\sim$}}}}
\def\gsim{\mathrel{\raise.3ex\hbox{$>$\kern-.75em\lower1ex\hbox{$\sim$}}}}

\nc{\non}{\nonumber}

\begin{document}

\begin{flushright}

 \mbox{\normalsize \rm CUMQ/HEP 162}\\
        \end{flushright}
\vskip 20pt

\title{\bf Higgs Phenomenology in Warped Extra-Dimensions with a 4th Generation}

\vskip 20pt

\author{Mariana Frank\footnote{mfrank@alcor.concordia.ca}}
\affiliation{Department of Physics, Concordia University\\
7141 Sherbrooke St. West, Montreal\\ Quebec, CANADA H4B 1R6
}
\author{
Beste Korutlu\footnote{beste.korutlu@gmail.com}}
\affiliation{Department of Physics, Concordia University\\
7141 Sherbrooke St. West, Montreal\\ Quebec, CANADA H4B 1R6
}
\author{Manuel Toharia\footnote{mtoharia@physics.concordia.ca}}
\affiliation{Department of Physics, Concordia University\\
7141 Sherbrooke St. West, Montreal\\ Quebec, CANADA H4B 1R6
}
\date{\today}

\begin{abstract}

We study a warped extra-dimension scenario where the Standard Model
fields lie in the bulk, with the addition of a fourth family of
fermions. We concentrate on the flavor structure of the Higgs  
couplings with fermions in the flavor anarchy ansatz. Even without a
fourth family, these couplings will be generically misaligned with
respect to the SM fermion mass matrices.
The presence of the fourth family typically enhances the misalignment
effects and we show that one should expect them to be 
highly non-symmetrical in the ${(34)}$ inter-generational mixing. 
The radiative corrections from the new fermions and their flavor
violating couplings to the Higgs affect negligibly known experimental
precision measurements such as the oblique parameters and $Z\to b
{\bar  b}$ or $Z \to \mu^+ \mu^-$. On the other hand, $\Delta F=1,2$
processes, mediated by tree-level Higgs exchange, as well as
radiative corrections to $b \to s \gamma$ and  $\mu \to e\gamma$ 
put some generic pressure on the allowed size of the flavor violating
couplings.    
But more importantly, these couplings will
alter the Higgs decay patterns as well as those of the new fermions,
and produce very interesting new signals associated to Higgs
phenomenology in high energy colliders. These might become very
important indirect signals for these type of models as they would be
present even when the KK mass scale is high and no heavy KK particle
is discovered.


\end{abstract}

\maketitle

\section{Introduction}


The Standard Model (SM) of particle physics has been remarkably
successful in explaining a wide range of low-energy phenomena and has
passed numerous experimental tests over the past few decades. The 
only ingredient of this model that has yet to be discovered is the
Higgs boson.  If the Higgs is discovered at the LHC, the problem of
its mass, which should receive quadratic corrections sensitive to
scales well above the electroweak scale (the hierarchy problem) still
remains. Warped extra dimensional models were introduced by Randall and
Sundrum (RS) \cite{RS} as an attempt to resolve this problem, by using
the extra-dimensional space-time warp factor to lower the natural scale
of particle masses. In the original model, all SM fields were
localized on the TeV brane, one could in principle generate higher
dimensional flavor violating operators, suppressed by only TeV
operators--a serious problem for phenomenology. To address this
issue one could invoke for example flavor symmetries \cite{Sundrum:2009gv}
but the most popular venue has been to allow fermions to propagate in
the bulk, which not only reduced the flavor problem, but provided a
compelling theory of flavor, in which hierarchies among fermion
masses and mixings arise naturally \cite{Davoudiasl, Grossman:1999ra}.
This model sheds light on the flavor puzzle as well:  the 5D Yukawa couplings
are  all ${\cal   O}(1)$ and with no definite flavor structure, and the fermion
masses and mixing angles depend on the amount of mixing of the
elementary fermions with the strongly coupled conformal field theory,
assumed to be small for the first two generations
\cite{Grossman:1999ra}. This implies that flavor 
violation in the SM is also suppressed by the same mixing factors -
the phenomenon that goes under the name of the RS-GIM mechanism
\cite{agashe}. However, in this case constraints from
the $\Delta S=2$ and $\Delta B=1$ processes still require the scale of
new physics (the KK scale) to be around $\sim 5$ TeV \cite{Agashe2site,
Gedalia:2009ws},
raising difficulties with observation of  this minimal scenario directly at
the LHC.


On the flavor side, recently, some possible deviations from the SM in $B$
meson \cite{Bphysics,Abazov:2011yk} and  $t$ quark physics \cite{newcdf} have been
reported, indicating perhaps difficulties with the standard
Cabibbo-Kobayashi-Maskawa (CKM) paradigm for quark mixing
\cite{Lunghi:2010gv}. The effects in $B$ physics can be explained by
various Beyond the Standard Model (BSM) scenarios, though the simplest
explanation seems to come from a simple extension of the Standard
Model to four generations, that is, by adding two new heavy quarks, a
heavy charge $+2/3$ quark ($t^\prime$) and charge $-1/3$ quark 
($b^\prime$). In more extensive versions of the model, the effects of
introducing extra leptons ($\tau^\prime$ and $\nu_{\tau}^\prime$),
needed for anomaly cancellation, are also studied. 

The addition of a fourth sequential generation of fermion doublets is
a natural extension of the SM (SM4). The model  restricts 
fourth-generation quark masses to be not too large to preserve
perturbativity \cite{marciano}.  Recently, SM4 have increased in
popularity as it was shown that the introduction of a fourth
generation does not conflict with electroweak precision observables
\cite{holdom}, as long as their mass differences are small
\cite{yanir}. Fourth generation fermions are required to have masses
greater than half the mass of the $Z^0$ boson to evade LEP limits on the
invisible $Z^0$ boson width.  There are many advantages of introducing
an extra family of fermions:  
\begin{itemize}
\item{These new fermions may trigger dynamical electroweak symmetry
  breaking \cite{marciano}  without a Higgs boson, and thus address
  the hierarchy problem.} 
\item{ A fourth generation softens the
current low Higgs mass bounds from electroweak precision observables
by allowing considerably higher values for the Higgs mass
\cite{flacher}.}
\item{ Gauge couplings can in principle be unified without invoking SUSY
\cite{hung}.}
\item{ A new family might cure certain problems in flavor physics,
  such as the CP-violation in $B_s$-mixing \cite{soni}.} 
\item {A fourth generation might solve problems related to
  baryogenesis, as an additional quark doublet could 
lead to a sizable increase of the measure of CP-violation \cite{hou}. }
\item {Such an extension of the SM would increase the strength of the
  phase transition  \cite{carena}.} 
\item {It appears that an even number of fermion generations is more
  natural from the string theory point of view \cite{Cvetic:2001nr}.}
\end {itemize}
New heavy fermions lead to new interesting effects due to their large Yukawa
couplings 
\cite{hung2}. 
Recent searches by the CDF Collaboration for direct
production of the fourth generation quarks, called $t^\prime$
and $b^\prime$, set the limits 
$m_{t^\prime} > 335$ GeV \cite{CDF} and $m_{b^\prime} > 385$ GeV
\cite{CDF2}, assuming Br$ \left(t^\prime \to W q(q = d, s, b)\right) = 100\%$
and Br$ (b^\prime \to Wt) = 100\%$ respectively. For the leptons
 $m_{\tau^\prime} > 100.8$ GeV,~$m_{\nu_{\tau}^\prime} >
90.3$ GeV~ (Dirac type),~$m_{\nu_{\tau}^\prime} > 80.5$ GeV (Majorana
type) \cite{particledata}. The limits on the low energy phenomenology due to fourth
generation fermions has been studied extensively
\cite{UTfit,Alok:2010zj}. 

While there have been many extensive studies of the SM4, there are few analyzes of BSM scenarios
with four generations (see however \cite{DePree:2009ed}).  The reason is that
the fourth generation
typically imposes severe restrictions on the models. In particular, there are
difficulties in incorporating a chiral fourth family scenario into any
Higgs doublet model, such as the MSSM \cite{Murdock:2008rx}. It was
initially shown that due to the large masses for the fourth generation
quarks and large Yukawa couplings, there are no values of
$\displaystyle \tan \beta=\frac{v_u}{v_d}>1$ for which the couplings
are perturbative to the Grand Unification Scale. (However, this
condition does not apply  to vector-like quarks \cite{Atre:2011ae}.)
Recently the MSSM with four generations has received some more attention
\cite{Cotta:2011ht},  as it was shown that for $\tan \beta \simeq 1$ the
model exhibits a strong first order phase transition
\cite{Fok:2008yg}.  

But  the four generation scenario can easily be incorporated
in models with warped extra dimensions, as in \cite{Burdman:2009ih},
where it can be argued that the fourth generation arises naturally. In
these models the Higgs particle can be thought of as a generic composite
state, and even being a condensate of some of the fourth
generation heavy quarks \cite{Burdman:2009ih,BarShalom:2010bh}, thus
providing a solution to the (little) hierarchy problem.   

An additional benefit of the extension of a fourth generation in
warped models, could be the inclusion of the fourth generation neutrino,
which 
may become a novel dark matter candidate \cite{Lee:2011jk}, typically
missing in minimal models (see however \cite{Agashe:2009ja} for
different approaches).


As mentioned earlier, KK particles could be just barely beyond the reach
of the LHC. Nevertheless there are implications of the warped
scenarios that could leave an imprint on lower energy physics. For
instance,  recently it was pointed out that 
warped extra-dimensional models introduce new flavor-violating
operators in the Higgs sector. In a composite Higgs sector with strong
dynamics, flavor changing neutral currents (FCNC) can arise at tree level,
generated by a misalignment between the Higgs Yukawa matrices and the
fermion mass matrices
\cite{Buchmuller:1985jz,Agashe:2009di,
  toharia1}. The  full set of operators responsible for
the misalignment has been thoroughly analyzed, showing that the  
effect is generically large and phenomenologically important
\cite{toharia1} and even could alter considerably the couplings of
Higgs to gluons \cite{Casagrande:2010si,toharia2}, affecting thus the main production
mechanism of the Higgs at hadron colliders.   

These flavor violating effects will be even more pronounced if the matter
sector is extended by extra fermionic generations. And for the Higgs
bosons, it is well known that the effects of a fourth generation are
quite spectacular in modifying the Higgs boson cross-section at hadron
colliders, which can be tested easily with Tevatron and early LHC data
within this or the next year. The Tevatron has published limits
on the Higgs boson cross-section in the fourth generation model,
excluding a wide range of Higgs boson masses \cite{HCDF}, and recently
the CMS collaboration carried out a similar study \cite{Chatrchyan:2011tz}.

As Higgs production can be modified within warped scenarios due to
flavor violating effects in the Higgs sector \cite{toharia2}, 
it may be possible to distinguish signals coming from a  fourth
generation model within the SM (SM4) with those coming from a fourth generation model
associated with a warped extra-dimension (or a composite scenario),
and, given the searches for the Higgs boson underway at the LHC,  such
an analysis is timely. The inclusion of the fourth generation will
also affect low-energy precision observables, as well as limits on
rare decays.  
In the lines of \cite {toharia1}, we propose to explore here the effect of
FCNC Higgs couplings with a fourth generation in a simple warped extra
dimensional model.


Our work is organized as follows. In the next section, 
Sec. \ref{sec:model}, we summarize the features of the warped
extra-dimensional models with fermions propagating in the bulk. We
analyze the flavor structure with fourth generational mixing in
Sec. \ref{sec:flavor}, giving both analytical expressions and numerical values.
We proceed to explore the 
phenomenology of the model in Sec. \ref{sec:pheno}. Restrictions due to
flavor-changing low energy observables, both at tree  and one-loop level, are
included here. In subsections, we investigate FCNC decays of the Higgs boson, as well as
collider signals for the fourth generation decaying into lighter fermions and
Higgs bosons. We summarize our findings and conclude in
Sec.~\ref{sec:conclusion}. In the Appendices we include some details
of our analytical evaluation.

\section{The model}
\label{sec:model}

For simplicity, we consider the simplest 5D warped extension of
the SM, in which we keep the SM local gauge groups and just extend the 
space-time by one warped extra dimension. There are bounds on the KK
scale coming from precision electroweak observables \cite{Agashe:2007mc} which
can
be addressed by extending the gauge group in order to obtain additional 
protection. Nevertheless the effects we are interested in lie in a
different sector of the scenario, namely the Higgs sector, and its
couplings with fermions. Our results can easily be extended to more
involved scenarios, but we feel it is best to show explicitly the effects in
the simplest scenario. Moreover precision electroweak constraints can
become milder with a heavier Higgs \cite{Casagrande:2008hr} and perhaps even if
the KK
scale is barely beyond LHC reach, one can observe its indirect effects in the
Higgs sector.

The spacetime we consider takes the usual Randall-Sundrum form~\cite{RS}:
\bea
ds^2
= \frac{R^2}{z^2}\! \Big(\eta_{\mu\nu} dx^\mu dx^\nu -dz^2\Big),
\label{RS}
\eea
with the UV (IR) branes localized at $z = R$ ($z = R^\prime$). We 
first focus  on a single family of down-type quarks $Q$, 
$D$. They contain the 4D SM $SU(2)_L$ doublet and singlet fermions
respectively with a 5D action 
\bea
\label{fermionaction} &&\hspace{-.5cm}
S_{\text{fermion}}\!=\!\int d^4x dz \sqrt{g} \Big[ {i \over 2}
\left(\bar{Q} \Gamma^A {\cal D}_A Q - {\cal D}_A \bar{ Q} \Gamma^A
Q\right)  
+ {c_{q} \over R} \bar{ Q} {Q} + (Q\rightarrow D),
\eea
where $c_{q}$ and $c_{d}$ are the 5D fermion mass coefficients.
We also consider a brane localized Higgs, and so the Yukawa
couplings in the Lagrangian are included in the action
\begin{equation}
S_{\text{brane}} = \int d^4x dz\, \delta(z- R') \left(\frac{R}{z}
\right)^4 H \left( {Y^{5D}_1} R \bar{{\cal Q}}_L {\cal D}_R +
{Y^{5D}_2} R\bar{{\cal Q}}_R {\cal D}_L + \text{h.c.} \right).
\label{braneaction}
\end{equation}

To obtain a chiral spectrum, we choose the following boundary conditions for $Q,
D$
\begin{eqnarray}
Q_L (+ +),\quad Q_R(- -), \quad D_L(- -), \quad D_R(+ +).
\end{eqnarray}
Then, only $Q_L$ and $D_R$ will have zero modes, with wavefunctions:
\begin{eqnarray}
q_L^0(z) &=& f(c_q)\frac{{R'}^{-\frac{1}{2}+c_q}}{ R^{2}} z^{2-c_q}, \\
d_R^0(z) &=& f(-c_d)\frac{{R'}^{-\frac{1}{2}-c_d}}{ R^{2}} z^{2+c_d},
\end{eqnarray}
where we have defined $f(c) \equiv
\sqrt{\frac{1-2c}{1-\epsilon^{1-2c}}}$ and the hierarchically small
parameter $\epsilon=R/ R'\approx 10^{-15}$, which is generally
referred to as the warp factor. Thus, if we choose $c_q (-c_d) > 1/2$,
then the zero modes wavefunctions are localized towards the UV brane; if
$c_q (-c_d) < 1/2$, they are localized towards the IR brane. The
wavefunctions of the fermion KK modes are all localized near the IR
brane. Note that the wavefunctions of the KK modes $Q_R$ and $D_L$
vanish at the IR brane due to their boundary conditions. The Yukawa
couplings of the Higgs with fermions (zero modes or heavy KK modes)
are set by the overlap integrals of the corresponding
wavefunctions. For a bulk Higgs localized near the IR brane,
the zero-zero-Higgs, zero-KK-Higgs, KK-KK-Higgs Yukawa couplings are
given approximately by
\begin{eqnarray}
Y_{d,00} &\sim& Y_* f(c_q) f(-c_d),\\
Y_{d, 0n} &\sim& Y_* f(c_q) \, \ \text{or}\ \, Y_* f(-c_d), \\
Y_{d, nm} &\sim& Y_*, 
\end{eqnarray}
where $Y_* = Y_d /\sqrt{R}$ is the ${\cal O}(1)$ dimensionless 5D Yukawa
coupling, and we ignored ${\cal O}(1)$ factors in the equations above. The
SM fermions are mostly zero mode fermions with some small amount of
mixing with KK mode fermions. Therefore, we can use the mass
insertion approximation to calculate the corrections to the masses and
Yukawa couplings of SM fermions. 
\begin{figure}[t]
 \center
 \includegraphics[height=3cm]{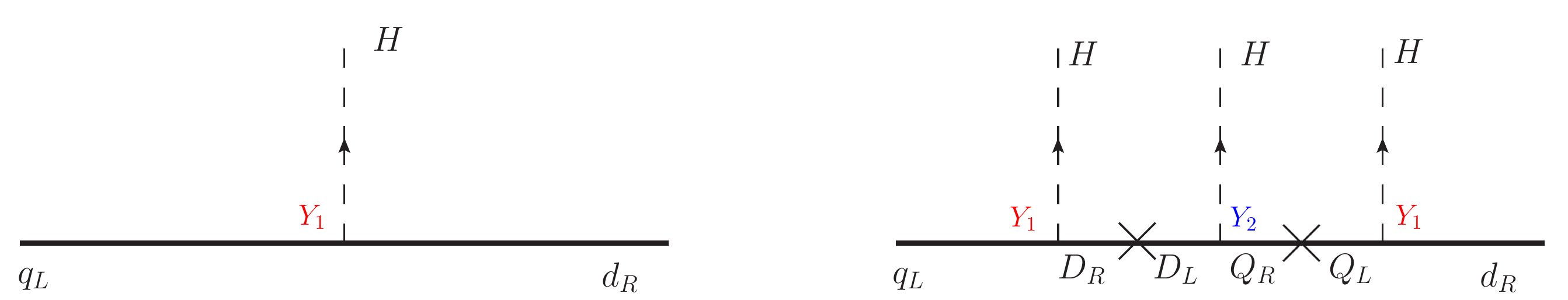}
 \caption{Correction to fermion mass and to physical
   Yukawa coupling (right diagram) after integrating out heavy
   vector-like fermions (the fermion KK modes here). In the right diagram, the
   correction to the mass happens when all the Higgs bosons acquire their
   VEV. The correction to the Yukawa coupling occurs when one of the
   Higgs remains physical; since there are three ways of doing this, mass
   correction and Yukawa correction are not the same, creating a shift
   between both terms (or a misalignment in flavor space). } 
\label{insertion}
 \end{figure}
This is shown in Fig. \ref{insertion} 
, where $q_L$, $d_R$ are zero modes of $SU(2)_L$ doublet and singlet
fermions respectively and $Q_L$, $Q_R$, $D_L$,
$D_R$ are KK mode fermions. One finds
\begin{eqnarray}
m^d_{SM} &\approx& 
f(c_q) Y_* f(-c_d)\ v_4 - f(c_q) \frac{Y_*^2 v_4^2}{M_{KK}^2} f(-c_d)
Y_*\ v_4, 
\end{eqnarray}
where $v_4$ is the Higgs vacuum expectation values (VEV) and we assume that all
KK fermion
masses are of the same order ($M_{KK}$).
The 4D effective Yukawa couplings of SM fermions can be calculated
using the same diagram, but the correction will be different. This is because in
the second diagram of Fig. \ref{insertion}, we have to set two external Higgs bosons $H$ to
their VEV
$v_4$ while the other one becomes the physical Higgs $h$, and there
are three different ways to do this. Thus we obtain the 4D Yukawa couplings
\begin{equation}
y^d_{SM} \approx f(c_q) Y_* f(-c_d)  - 3 f(c_q) \frac{Y_*^2v_4^2}{M_{KK}^2}
f(-c_d) Y_*.
\end{equation}
We see that the SM fermion masses and the 4D Yukawa couplings are not
universally proportional; indeed there is a shift with respect to the
SM prediction of  $m^d_{SM}=y^d_{SM}v_4$.

This shift, or misalignment, defined as $\Delta^d=m^d - Y^d v_4$ can
be carefully calculated perturbatively including ${\cal O}(1)$
factors. It is found to be \cite{toharia1}
\begin{equation}
\label{deltabrane}
\Delta^d_{_1}=\frac{2}{3} m_d Y^{5D}_1  (Y^{5D}_2)^* v_4^2 R'^2 =
\frac{2}{3} |m_d|^2 m_d R'^2
\left(\frac{Y_2^{5D}}{Y_1^{5D}}\right)^* \frac{1}{f(c_q)^2f(-c_d)^2}.
\end{equation}
Note the presence of the independent couplings
$Y_2^{5D}$ which are not necessary for generating fermion masses. 
It is technically possible to set their values as small as necessary
and suppress the misalignment. Nevertheless this seems to go against the main
philosophy of our approach which assumes that the value of all
dimensionless 5D parameters is of order one. Moreover in the case
where the Higgs is a bulk scalar field we have $Y_1=Y_2$, which is the
simplifying assumption we will make for our numerical computations.

There is another contribution to the misalignment which can be also calculated
and is given by \cite{toharia1}
\begin{eqnarray}
\Delta^d_{2} = m_d |m_d|^2 R'^2 \left[K(c_q) + K(-c_d)\right],
\end{eqnarray}
with 
\begin{equation} K(c) \equiv \frac{1}{1-2c}\left[
-\frac{1}{\epsilon^{2c-1}-1}+ \frac{\epsilon^{2c-1}-\epsilon^2}
{(\epsilon^{2c-1}-1)(3-2c)}+\frac{\epsilon^{1-2c}-\epsilon^2}
{(1+2c)(\epsilon^{2c-1}-1)}\right]. 
\label{Kc}
\end{equation}
One can see that $\Delta^d_{1}$ and $\Delta^d_{2}$ can be of the same
parametric order only for IR localized fermions (heavy quarks), but  will be
quite
suppressed for light quarks.

\section{Flavor Structure with four families}
\label{sec:flavor}

We now proceed to add to the scenario the remaining families of quarks
and leptons, including a new fourth generation. This will of course
create a richer structure of flavor, not only in the Higgs sector, but
in the electroweak sector, where the flavor changing charged current
mediated by $W$ bosons will now contain new vertices with the addition of $t'$
and $b'$.

The fermion wavefunctions evaluated at the TeV brane ($f(c)$) are now
promoted to diagonal matrices $\hat{F}_{q} =
\text{diag}[f({c_{q_i}})]$ and $\hat{F}_{d} =
\text{diag}[f({c_{d_i}})]$. Small differences in the $c's$ will
produce large hierarchies in the values of $f(c)$ (i.e. geographical
fermion localization in the extra dimension), and so the
matrices $\hat{F}_{q,d}$ are highly hierarchical, leading to mass
hierarchies and small mixing angles.

\subsection{The quark mixing matrix $V_{CKM4}$}

\begin{figure}[t]
 \center
 \includegraphics[width=9cm,height=5cm]{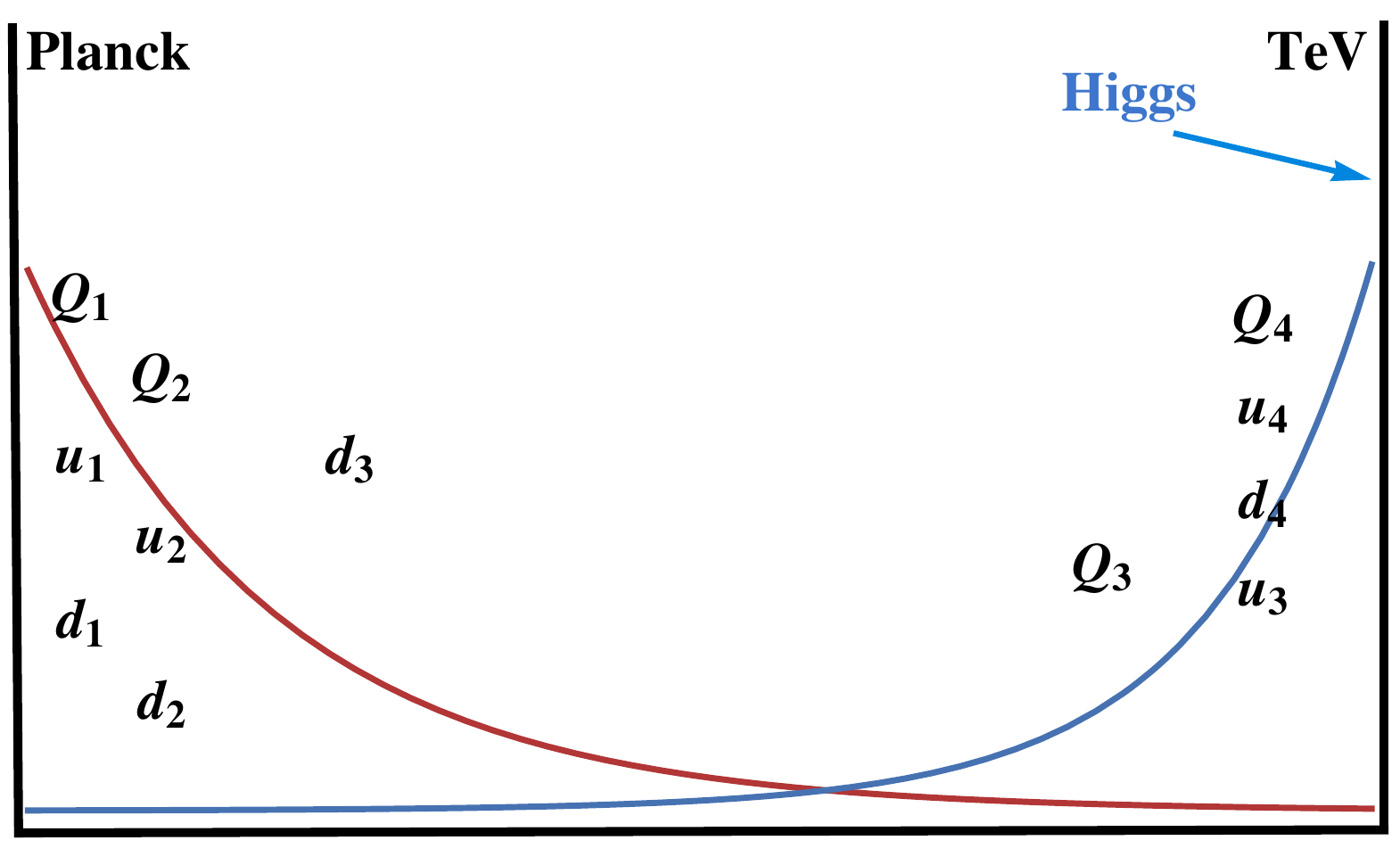}
 \caption{Typical geographic location of quarks in RS-4GEN (RS with a
   fourth family) such that large quark
   mass hierarchies and small mixing angles are generic. The Higgs boson and
   the heavier fermions (top and fourth generation quarks and charged
   leptons) are localized near the  TeV brane, whereas light fermions
   are localized towards the Planck brane.} 
\label{artistic}
 \end{figure}

The mass matrices are
\bea
{\bf m_u} &=& F_Q\ Y_u\ F_u,\\\
{\bf m_d} &=& F_Q\ Y_d\ F_d, 
\eea
where $F_Q$, $F_u$ and $F_d$ are $4\times4$ diagonal matrices whose
entries are given by the values at the IR brane of the corresponding zero-mode
wave
functions:
\bea
F_Q=\left(\begin{tabular}{cccc}
$f_{Q_1} $ & & \\
&$f_{Q_2}$  & \\
& &$f_{Q_3}$\\
& & & $f_{Q_4}$\\
\end{tabular} \right),\ \ \   
F_u=\left(\begin{tabular}{cccc}
$f_{u_1} $ & & \\
&$f_{u_2}$  & \\
& &$f_{u_3}$\\
& & &$f_{u_4}$\\
\end{tabular} \right),\ \ \   
F_d=\left(\begin{tabular}{cccc}
$f_{d_1} $ & & \\
&$f_{d_2}$  & \\
& &$f_{d_3}$\\
& & &$f_{d_4}$\\
\end{tabular} \right).
\eea
The matrices $Y_u$ and $Y_d$ are the 5-dimensional Yukawa
couplings, i.e. general $4\times4$ complex matrices. Because most of
the entries in the diagonal matrices $F_q$ are naturally
hierarchical (for UV-localized fermions), the physical fermion mass
matrices $m_u$ and $m_d$ will 
inherit their hierarchical structure independently of the nature of
the true 5D Yukawa couplings $Y_u$ and $Y_d$, which can therefore contain all
of its entries with similar size (of order 1) and have no definite
flavor structure. This is the main idea behind scenarios of so-called
flavor anarchy, which we consider also here, but applied to a
four-family scenario. 
The introduction of the fourth family is simply realized by
assuming that the new fermions are localized near the TeV brane, like
the top quark, and therefore will be naturally heavy. Mixing angles
should typically be small except among the heavy fermions where
large mixings could be possible.

To diagonalize the mass matrices we use 
\bea
\hspace{4cm}  U_{Q_u}\ {\bf  m_u}\ W_u^\dagger &=& {\bf m_u^{diag}}, 
\label{UW}\\ 
\hspace{4cm}  U_{Q_d}\ {\bf  m_d}\ W_d^\dagger &=& {\bf m_d^{diag}} . 
\eea
One can in fact obtain a relatively simple formulation of the rotation
matrices $U_{Q_u}$, $U_{Q_d}$, $W_u$ and $W_{d}$ by expanding their
entries in powers of ratios $f_i/f_j$, where $i<j$ and with $i=1,2$
and $j=1,2,3,4$.
To proceed, we first define our notation. If ${\bf A}$ is an
$n\times n$ matrix, then ${[\bf A]}_{ij}$ represents its $\{ij\}$
first order minor, i.e. the determinant of the $(n-1) \times (n-1)$
submatrix obtained by removing row $i$ and column $j$ from ${\bf A}$. We
 also use the notation ${\bf[A]}_{ij,\alpha\beta}$ to represent
the $\{ij, \alpha \beta\}$ second order minor of ${\bf A}$, i.e. the
determinant of the  $(n-2) \times (n-2)$ submatrix obtained by
removing rows $i$ and $\alpha$, and columns $j$ and $\beta$ from the
matrix ${\bf A}$. 
Keeping only the leading terms, we obtain (see \cite{Casagrande:2008hr}
 for the three family case):
\bea
U_{Q_u} &=& \left(
\begin{tabular}{cccc}
$1\ $ & $\vphantom{\frac{\int_\int}{\int_{\int_\int}}} \displaystyle
  \frac{[Y_u]_{{}_{21}}}{[Y_u]_{{}_{11}}}\ \frac{f_{Q_1}}{f_{Q_2}}\ \ $
&  ${\cal U}^{Q_u}_{13}\ \ $&  ${\cal U}^{Q_u}_{14}\ \ $
\\ 
$\vphantom{\frac{\int_\int}{\int_{\int_\int}}} -\displaystyle
  \frac{[Y_u]^{{}^*}_{{}_{21}}}{[Y_u]^{{}^*}_{{}_{11}}}\
\frac{f_{Q_1}}{f_{Q_2}}\ \ $
  & \ 1 & 
 ${\cal U}^{Q_u}_{23}\ \ $&  ${\cal U}^{Q_u}_{24}\ \ $
\\
$\ \ \vphantom{\frac{\int_\int}{\int_{\int_\int}}}
  \displaystyle\frac{[Y_u]^{{}^*}_{{}_{31}}}{[Y_u]^{{}^*}_{{}_{11}}}\
\frac{f_{Q_1}}{f_{Q_3}}\ $ & 
  $\vphantom{\frac{\int_\int}{\int_{\int_\int}}}
-\displaystyle\frac{[Y_u]^{{}^*}_{{}_{11,32}}}{[Y_u]^{{}^*}_{{}_{11,22}}}\    
\frac{f_{Q_2}}{f_{Q_3}}\ \ $   & $c_{Q_u}$ & $s_{Q_u}$\\  
$\vphantom{\frac{\int_\int}{\int_{\int_\int}}}
-  \displaystyle\frac{[Y_u]^{{}^*}_{{}_{41}}}{[Y_u]^{{}^*}_{{}_{11}}}\
\frac{f_{Q_1}}{f_{Q_4}}\ $  &   
 $\ \ \vphantom{\frac{\int_\int}{\int_{\int_\int}}}
  \displaystyle\frac{[Y_u]^{{}^*}_{{}_{11,42}}}{[Y_u]^{{}^*}_{{}_{11,22}}}\   
\frac{f_{Q_2}}{f_{Q_4}}\ \ $
  & $-s^*_{Q_u}$& $c^*_{Q_u}$\\
\end{tabular} \right),
\label{UQ}
\eea 


\bea
U_{Q_d} &=& \left(
\begin{tabular}{cccc}
$1\ $ & $\vphantom{\frac{\int_\int}{\int_{\int_\int}}} \displaystyle
  \frac{[Y_d]_{{}_{21}}}{[Y_d]_{{}_{11}}}\ \frac{f_{Q_1}}{f_{Q_2}}\ \ $
& ${\cal U}^{Q_d}_{13}\ \ $&  ${\cal U}^{Q_d}_{14}\ \ $\\ 
$\vphantom{\frac{\int_\int}{\int_{\int_\int}}} -\displaystyle
  \frac{[Y_d]^{{}^*}_{{}_{21}}}{[Y_d]^{{}^*}_{{}_{11}}}\
\frac{f_{Q_1}}{f_{Q_2}}\ \ $
  & \ 1 &${\cal U}^{Q_d}_{23}\ \ $&  ${\cal U}^{Q_d}_{24}\ \ $
  \\
$\ \ \vphantom{\frac{\int_\int}{\int_{\int_\int}}}
  \displaystyle\frac{[Y_d]^{{}^*}_{{}_{31}}}{[Y_d]^{{}^*}_{{}_{11}}}\
\frac{f_{Q_1}}{f_{Q_3}}\ $ & 
  $\vphantom{\frac{\int_\int}{\int_{\int_\int}}}
-\displaystyle\frac{[Y_d]^{{}^*}_{{}_{11,32}}}{[Y_d]^{{}^*}_{{}_{11,22}}}\    
\frac{f_{Q_2}}{f_{Q_3}}\ \ $   & $c_{Q_d}$ & $s_{Q_d}$\\  
$\vphantom{\frac{\int_\int}{\int_{\int_\int}}}
-  \displaystyle\frac{[Y_d]^{{}^*}_{{}_{41}}}{[Y_d]^{{}^*}_{{}_{11}}}\
\frac{f_{Q_1}}{f_{Q_4}}\ $  &   
 $\ \ \vphantom{\frac{\int_\int}{\int_{\int_\int}}}
  \displaystyle\frac{[Y_d]^{{}^*}_{{}_{11,42}}}{[Y_d]^{{}^*}_{{}_{11,22}}}\   
\frac{f_{Q_2}}{f_{Q_4}}\ \ $
  & $-s^*_{Q_d}$& $c^*_{Q_d}$\\
\end{tabular} \right),
\eea 
where, in particular, we have
\bea
{\cal U}^{Q_d}_{23}&=& c_{Q_d} \frac{f_{Q_2}}{f_{Q_3}}\ 
\frac{[Y_d]_{{}_{11,32}}}{[Y_d]_{{}_{11,22}}}\ +  s^*_{Q_d}
\frac{f_{Q_2}}{f_{Q_4}}\ 
\frac{[Y_d]_{{}_{11,42}}}{[Y_d]_{{}_{11,22}}},\\
{\cal U}^{Q_d}_{13}&=&c_{Q_d} \frac{f_{Q_1}}{f_{Q_3}}\ 
\frac{[Y_d]_{{}_{21,32}}}{[Y_d]_{{}_{11,22}}}\ +  s^*_{Q_d}
\frac{f_{Q_1}}{f_{Q_4}}\ 
\frac{[Y_d]_{{}_{21,42}}}{[Y_d]_{{}_{11,22}}}, 
\eea
where we used properties of the minors. These are needed to compute the $V_{CKM}$ elements $V_{cb}$ and $V_{ub}$.
Due to the mass hierarchy $m_b\ll m_{b'}$, we also have the simple expansions:
\bea
c_{Q_d}= v_4\ f_{Q_4} f_{d_4} |Y^d_{44}|/m_{b'}\ \ \ \  {\rm and}\ \  \ \
s^*_{Q_d}=
v_4\ f_{Q_3} f_{d_4} {Y^d_{34}}^*/m_{b'}\ e^{i\arg{(Y^d_{44})}}.
\eea 
Since 
\bea
V_{CKM}=U_{Q_u}^+U_{Q_d} ,
\eea
 we can find expressions
for $V_{us}$, $V_{cb}$ and $V_{ub}$: 
\bea
\label{vus}
V_{us}=\frac{f_{Q_1}}{f_{Q_2}}\left(
\frac{[Y_d]_{{}_{21}}}{[Y_d]_{{}_{11}}}-
\frac{[Y_u]_{{}_{21}}}{[Y_u]_{{}_{11}}}\right) , 
\eea
and
\bea
\label{vcb}
V_{cb}=c_{Q_d}
\frac{f_{Q_2}}{f_{Q_3}}\ \left(\frac{[Y_d]_{{}_{11,32}}}{[Y_d]_{{}_{11,22}}}\ -
 \frac{[Y_u]_{{}_{11,32}}}{[Y_u]_{{}_{11,22}}}\
\right)\
+\  s^*_{Q_d} \frac{f_{Q_2}}{f_{Q_4}}
\left(\frac{[Y_d]_{{}_{11,42}}}{[Y_d]_{{}_{11,22}}}\
- \frac{[Y_u]_{{}_{11,42}}}{[Y_u]_{{}_{11,22}}}\
\right), 
\eea
and
\bea
\label{vub}
V_{ub}&&=c_{Q_d}\frac{f_{Q_1}}{f_{Q_3}} 
\left(
\frac{[Y_u]_{{}_{31}}}{[Y_u]_{{}_{11}}} +
\frac{[Y_d]_{{}_{21,32}}}{[Y_d]_{{}_{11,22}}} -
\frac{[Y_u]_{{}_{21}}}{[Y_u]_{{}_{11}}}
\frac{[Y_d]_{{}_{11,32}}}{[Y_d]_{{}_{11,22}}}
\right)\!
\non\\
 &&  
+  s^*_{Q_d} \frac{f_{Q_1}}{f_{Q_4}} 
\left(
\frac{[Y_u]_{{}_{41}}}{[Y_u]_{{}_{11}}} +
\frac{[Y_d]_{{}_{21,42}}}{[Y_d]_{{}_{11,22}}} -
\frac{[Y_u]_{{}_{21}}}{[Y_u]_{{}_{11}}}
\frac{[Y_d]_{{}_{11,42}}}{[Y_d]_{{}_{11,22}}}
\right).\ \ \   
\eea
It is clear that if the 5D Yukawa matrix elements are all of order 1, then
the observed hierarchies among the CKM elements can still be explained by
hierarchies among the $f_i$ parameters. The explicit dependence on the
5D Yukawa couplings gives a more precise prediction for the mixing
angles, which will be quite useful when looking for phenomenologically
viable points in parameter space. The results of such a scan are
presented in the next subsection.

\subsection{Tree level Higgs FCNC couplings}

We now extend the one-family results presented in section
\ref{sec:model} to the case of four generations. To leading order in
Yukawa couplings, the SM fermion mass matrix is 
\begin{equation}
\hat{m}^d = \hat{F}_Q \hat{Y}_{1}^{5D} \hat{F}_d\ \ {v_4},
\end{equation}
where $\hat{}$ means a $4 \times 4$ matrix in flavor space. The
misalignment in flavor space between the fermion mass matrix and the
Yukawa coupling matrix is defined as   
\bea
\hat{\Delta}^d = \hat{m}^d - v_4\ \hat{y}_4^d,
\eea
where $\hat{y}_4^d$ is the 4D effective coupling matrix between the physical
scalar Higgs and the quarks.

Similarly to the one family case, the misalignment can be separated into two components,
$\hat{\Delta}^d_1 + \hat{\Delta}^d_2$, with (see \cite{toharia1})
\bea
\hat{\Delta}^d_{1}&=&
\frac{2}{3}\ \hat{m}^{d} \frac{1}{\hat{F}_d}
  (\hat{Y}^{5D}_2)^\dagger \frac{1}{\hat{F}_Q} \hat{m}^{d}
  \ \left({v_4^3 R'^2}\right),
\label{threegenresult12}
\eea
and
\begin{equation}
\label{threegenresult2}
\hat{\Delta}^d_{2} = \hat{m}^d \left(
\hat{m}^{d\dagger} \hat{K}(c_q) + \hat{K}(-c_d) \hat{m}^{d\dagger}
\right)\hat{m}^d\  R'^2
\end{equation}
The crucial observation
is that $\hat{m}^d $ and $\hat{\Delta}^d$ are generally not aligned in
flavor space. Thus when we diagonalize the quark mass matrix with a
bi-unitary transformation $\hat{m}^d \rightarrow U^\dagger_{Q_L}
\hat{m}^d W_{d}$, the Yukawa couplings will not be diagonal. To be
more specific, in models of flavor anarchy,
we have
\begin{equation}\label{fratio}
(U_{Q_d},W_d)_{i,j} \sim
\frac{f_{Q_i,d_i}}{f_{Q_j,d_j}} \qquad \text{for}
\quad {i < j}.
\end{equation}
Then the off-diagonal Yukawa coupling will be dominated by 
\begin{eqnarray}
\label{offcoup}
\hat{Y}^{\text{off}}_{ij}& = & -(U^\dagger_{d_L}
\hat{\Delta}^d W_{d_R})_{ij}\frac{1}{v_4}\sim 
\frac{2}{3} f_{Q_i}\bar{Y}^3 f_{d_j} v_4^2 R'^2,
\end{eqnarray}
where $\bar{Y}$ is the typical value of the dimensionless 5D Yukawa coupling.

Since the Higgs couplings will now contain off-diagonal entries, we
must choose a convenient parametrization for them. A common choice is
to normalize the couplings with the fermion masses and write the Higgs
Yukawa couplings as{\footnote{This is a particular realization of the Cheng-Sher {\it Ansatz} \cite{Cheng:1987rs}.}}
\begin{equation}
\label{aijdef}
{\cal{L}}_{HFV} = a^d_{ij}\sqrt{\frac{m^d_i m^d_j}{v_4^2}} H
\bar{d}_L^i d_R^j + h.c. + (d \leftrightarrow u).
\end{equation}

\subsubsection{Analytical estimates of Higgs FCNC couplings in Flavor Anarchy}

In this section, following the same procedure as in \cite{toharia1}, we
estimate the off-diagonal couplings of Higgs boson to 
SM fermions 
and then we do a numerical scan over anarchical Yukawa
couplings to support our estimates.

We use Eqs.~(\ref{fratio}) and (\ref{offcoup}) to estimate the
sizes of $a^{u,d}_{ij}$. For example, we have
\begin{eqnarray} 
a^d_{12} &\sim& \frac{2}{3} f_{Q_1} \bar{Y}^3 f_{d_2} v_4^2 R'^2
\sqrt{\frac{v_4^2}{m_s m_d}} 
\sim \frac{2}{3} \lambda \bar{Y}^2 v_4^2 R'^2
\sqrt{\frac{m_s}{m_d}},
\end{eqnarray}
where $\lambda \approx 0.22$ is the Wolfenstein parameter, and we
used $f_{q_1}/f_{q_2} \sim (U_{d_{L}})_{12}\sim (V_{CKM})_{12}\sim
\lambda$. We can find the other $a^{u,d}_{ij}$ in similar fashion.
We obtain:
\begin{eqnarray}
a^d_{ij} \sim \delta_{ij}  -\frac{2}{3}\bar{Y}^2 v_4^2
R'^2\left(\begin{array}{cccc} 1 &
\lambda\sqrt{\frac{m_s}{m_d}} & \lambda^3 \sqrt{\frac{m_b}{m_d}}& \lambda^3
\sqrt{\frac{m_{b'}}{m_d}}\\
\frac{1}{\lambda}\sqrt{\frac{m_d}{m_s}} & 4 & \lambda^2
\sqrt{\frac{m_b}{m_s}} & \lambda^2
\sqrt{\frac{m_{b'}}{m_s}}\\ 
\frac{1}{\lambda^3}\sqrt{\frac{m_d}{m_b}}&
\frac{1}{\lambda^2}\sqrt{\frac{m_s}{m_b}} & 12  & \sqrt{\frac{m_{b'}}{m_b}}\\
\frac{1}{\lambda^3}\sqrt{\frac{m_d}{m_{b'}}}
&\frac{1}{\lambda^2}\sqrt{\frac{m_s}{m_{b'}}} & \sqrt{\frac{m_{b}}{m_{b'}}}  &
12 
\end{array} \right),
\label{adest}
\end{eqnarray}
\begin{eqnarray}
a^u_{ij} \sim \delta_{ij} -\frac{2}{3}\bar{Y}^2 v_4^2
R'^2\left(\begin{array}{cccc} 1 &
\lambda\sqrt{\frac{m_c}{m_u}} & \lambda^3 \sqrt{\frac{v_4^2}{m_t m_u}}&
\lambda^3 \sqrt{\frac{v_4^2}{m_{t'} m_u}} \\
\frac{1}{\lambda}\sqrt{\frac{m_u}{m_c}} & 4 & \lambda^2
\sqrt{\frac{v^2_4}{m_tm_c}}&  \lambda^2
\sqrt{\frac{v^2_4}{m_{t'}m_c}} \\ 
\frac{1}{\lambda^3}\sqrt{\frac{m_u}{m_t}}
& \frac{1}{\lambda^2}\sqrt{\frac{m_c}{m_t}} & 16 &
\sqrt{\frac{m_{t'}m_t}{v_4^2}}\\
\frac{1}{\lambda^3}\sqrt{\frac{m_u}{m_{t'}}} &
\frac{1}{\lambda^2}\sqrt{\frac{m_c}{m_{t'}}} &
\sqrt{\frac{v_4^2}{m_{t'}m_t}}& 16 
\end{array} \right).
\label{auest}
\end{eqnarray}
The effect clearly decouples since it depends on $R'^2 \sim
\frac{1}{M_{KK}^2}$. Taking the typical Yukawa size $\bar{Y}=2$  and $1/R'=1500$
GeV, and using the known SM masses evaluated at the KK scale, along
with $m_{t^\prime}=400$ GeV and $m_{b^\prime}=350$ GeV, one can obtain the
typical values of these couplings:  
\begin{eqnarray}
a^d_{ij} \sim
 \left(\begin{array}{cccc} 0.96 &0.03 &0.01 &0.14\\
0.04 &0.86 &0.01 &0.15\\
0.13 &0.19 &0.57 &0.45\\
0.01 &0.007 &0.003 &0.57
\end{array} \right),
\label{aijdestimate}
\end{eqnarray}

\begin{eqnarray}
a^u_{ij} \sim
 \left(\begin{array}{cccc} 
0.96 &0.16 &0.15 &0.09\\
0.008 &0.86 &0.04 &0.02\\
0.01 &0.04 &0.42 &0.05\\
0.007 &0.03 &0.003 &0.42
\end{array} \right).
\label{aijuestimate}
\end{eqnarray}
Note that the results presented here are just estimates for the
size of $a^{u,d}_{ij}$, which come without sign or phases. However, we
observe that for the third and fourth generation quarks, the
corrections to the diagonal Yukawa couplings are always negative
(suppressions) if $Y_1 = Y_2$ and are larger than the previous
estimates. This point was argued in \cite{toharia1} and we address it
again the next subsection for completeness.

 An interesting feature of these matrices is the asymmetry of
 $a^d_{ij}$ in the $ b_L b_R^\prime$ and $b_L^\prime b_R$ entries, asymmetry
 not shared by the up-quark matrix $a^u_{ij}$. These would produce an
 asymmetry in the decays, as well as in the shift of the the vertex
 functions $g^b_L$, $g^b_R$ for $Z \to b \bar{b}$. This asymmetry will
 be typical to the $(34-43)$ entries and thus non-universal. We expect
 the same feature in the charged lepton mass matrix.

\subsubsection{Numerical results for Higgs FCNC couplings}
\label{numerical}

In order to obtain a better prediction of the typical size of the
off-diagonal Yukawa couplings, and to compare with the previous
estimates we perform a scan in parameter space. The results
should be in general consistent with the rough estimates of
Eqs.~(\ref{aijdestimate}) and
(\ref{aijuestimate}). Some differences observed can nevertheless be explained,
 (see also \cite{toharia1}) so that one can still be confident in the
generic size of the flavor violating couplings predicted in the flavor
anarchy paradigm in RS type scenarios with four generations.

We proceed as follows:
\bit
  \item We fix $m_{t'}=400$ GeV and $m_{b'}=350$ GeV as well as SM quark
    masses at the KK scale, taken to be $m_t=140$ GeV, $m_b=2.2$ GeV,
    $m_c=0.55$ GeV,  $m_s=5\times 10^{-2}$ GeV,  $m_u=1.5\times 10^{-3}$ GeV,
    $m_d=3.0\times 10^{-3}$ GeV. We take the KK
    scale as ${R'}^{-1}=1500$ GeV.   
  \item Then we generate random complex entries for $Y_u$ and $Y_d$,
    such that $|Y_i|\in [0.3,3.5]$. We also generate random
    $f_{Q_4}$ such that $f_{Q_4} \sim {\cal O} (1)$.  
  \item
  We then obtain $f_{Q_3}$ from $|V_{ub}|/|V_{us}|/|V_{cb}|$,
  $\ f_{Q_2}$  from $|V_{ub}|/|V_{us}|\ $  and $\ f_{Q_1}$ from
  $|V_{us}|$ (see Eqs.~(\ref{vus}), (\ref{vcb}) and (\ref{vub})).
  \item We then obtain the right-handed down quark entries
    $f_{d_4}$ from $m_{b^\prime}$. 
  \item Similarly for the up right-handed matrix entries, we obtain
    $f_{u_1}$, $f_{u_2}$,  $f_{d_1}$,  $f_{d_2}$ and  $f_{d_3}$ from
    $m_u, m_c, m_d,m_s$ and $m_b$. We also obtain  $f_{u_3}$ and
    $f_{u_4}$ from  $m_t$ and $m_{t^\prime}$.
  \item Finally we check that the generated $Y_u$ and $Y_d$ along
    with the obtained $\hat{F}_q$, $\hat{F}_u$ and $\hat{F}_d$ do
    indeed produce the observed masses and mixings of the SM. If so we
    keep the point in parameter space and continue until we obtain
    1000 points which satisfy all constraints.  
    \item For each acceptable point, we use
      Eqs.~(\ref{threegenresult12}) and (\ref{threegenresult2}) to
      compute the flavor  violating Higgs Yukawa couplings,
      parametrized by the $a_{ij}$'s as defined in Eq.~(\ref{aijdef}). 
\eit

We present the results of the scan as follows: we give the 25\% quantile
and the 75\% quantile of the obtained couplings. This means that 50\%
of our acceptable points contain a coupling in between the quoted
values. Also it means that 25\% of the generated points predict higher
values than the range quoted, while 25\% of the points predict lower
values than the range quoted. \\

We find the following ranges for $a^d_{ij},\  a^u_{ij}$ matrix couplings
\begin{eqnarray}
a^d_{ij} \sim \left(\begin{array}{cccc} 0.919-0.987 &
0.025- 0.081 & 0.011-0.044 & 0.130-0.532 \\
0.049-0.148 & 0.827-0.934 & 0.0.017-0.059&0.249-0.934\\ 
0.140-0.470&
0.142-0.446 & 0.620-0.819  &0.873-2.508\\
0.018-0.061 & 0.017-0.058 & 0.008-0.120 & 0.375-0.643 
\end{array} \right),
\label{adscan}
\end{eqnarray}

\begin{eqnarray}
a^u_{ij} \sim \left(\begin{array}{cccc} 0.927-1.000 &
0.089-0.364 & 0.091-0.410& 0.139-0.612\\
0.015-0.052 & 0.816-0.949&0.065-0.197& 0.092-0.300\\ 
0.019-0.068
&0.071-0.236 & 0.545-0.772 & 0.127-0.343\\
0.0167-0.062 & 0.060-0.191 & 0.064-0.168 & 0.403-0.651
\end{array} \right),
\label{auscan}
\end{eqnarray}

to be compared with the rough estimates Eqs.~(\ref{aijdestimate}) and
(\ref{aijuestimate}).

\subsubsection{Cumulative effect on diagonal Yukawa couplings when $Y_1 = Y_2$}
\label{3rdgeneration}

We observe that the rough estimates are slightly smaller than the
results of the scan, specially for the third and fourth generation couplings.
This was already pointed out in \cite{toharia1} for the three
generation case. The argument  given is that due to the
presence of a fourth generation some of the coefficients will be
different and typically the cumulative effect will be larger.

We assume that $Y_1=Y_2$\footnote{This is an important choice, and without it no
extra enhancements should appear. Nevertheless this choice is 
natural if the Higgs boson is to be considered as a highly localized 5D
scalar field, and then 5D Lorentz invariance imposes $Y_1=Y_2$.} and
consider for example the element ${(33)}$ of the Yukawa coupling in the
up quark sector, i.e.  
\bea
a_{tt}-1&=&-\frac{2{R'}^2}{3m_t} \left[ U^\dagger_{Q_u}
\hat{m}^u\frac{1}{\hat{F}_u^2}
\hat{m}^{u\dagger}\frac{1}{\hat{F}_Q^2} \hat{m}^u W_{u}
\right]_{33} \non\\
&=&-\frac{2{R'}^2}{3m_t}
\left( m_u^{diag}\right)_{33}
\left(W^\dagger_{u}\frac{1}{\hat{F}_u^2} W_{u}\right)_{3j}
\left(m_u^{diag}\right)_{jj}
\left(U^\dagger_{Q_u}\frac{1}{\hat{F}_Q^2} U_{Q_u}\right)_{j3}
\left(m_u^{diag}\right)_{33}.
\eea
First let's look at the contribution to $a_{tt}$ when the $j$ index
is equal to 3 (i.e. in the middle mass matrix $m_u^{diag}$ is
$m_t$). In this case, there will be 16 terms in phase, each proportional to
$-\frac{2{R'}^2 \bar{Y}^2 v_4^2}{3}  $, and it is important to
realize that every one of them will be real and negative, because
$(W^\dagger_{u}\frac{1}{\hat{F}_u^2} W_{u})_{33}\geq0$.  When
$j=2$ $(m_u^{diag}=m_c)$ there will be 2 terms $\sim\frac{2{R'}^2
\bar{Y}^2 v_4^2}{3}$ but every one of them will have generically a random
complex phase
 (the 14 remaining terms are much
smaller).  For $j=1$ $(m_u^{diag}=m_u)$ there is only one term   $\sim\frac{2{R'}^2
  \bar{Y}^2 v_4^2}{3}$ contributing, with the rest 15 terms being
again suppressed.  So, summing, the dominant contribution
to $a_{tt}$  will consist of 19 terms, 16 of which are negative and
the rest 3 have random complex phases. Generically each of these
terms are of the same size $\sim \frac{2{R'}^2 \bar{Y}^2 v^2}{3}$ so from
a statistical argument, $a_{tt}-1$ should receive a negative
contribution $\sim-16 \left( \frac{2{R'}^2 \bar{Y}^2 v^2}{3}\right)$.
This cumulative effect is confirmed by the numerical scan.\\

One can perform the same analysis for the rest of elements of the
Yukawa matrix, including the off diagonal ones, and realize that
typically there are a number of aligned terms in each case which
enhances the naive estimate by an ${\cal O}(1)$ factor (which can be
estimated also). This fact gives us confidence that both our scan and
our estimates are consistent and that our numerical results predict
correctly in this scenario  the generic size of the flavor violating couplings in the
Higgs sector.

\subsubsection{Higgs FCNC couplings in the lepton sector}
\label{leptons}

We proceed in a similar fashion to evaluate  Higgs
flavor violation in the lepton sector. The difficulty with the lepton sector is
that mixing matrices are not well-established here. The neutrinos can
be either Dirac or Majorana, the charged lepton mixing matrix (PMNS)
is not as well established as the CKM matrix, and  there  
are several mechanisms to explain  the large
mixing angles and light masses  for the neutrinos
(see for example \cite{Perez:2008ee,Agashe:2009tu}). For all cases,
the Lagrangian can then be parametrized  as: 
\begin{equation}
{\cal{L}}_{HFV} = a^l_{ij}\sqrt{\frac{m^l_i m^l_j}{v_4^2}} H
\bar{L}^i e^j + h.c.
\end{equation}
Following \cite{Agashe:2009tu,toharia1}, we analyze two types of scenarios. Depending
on the neutrino model, the left-handed charged lepton profiles
can be either hierarchical  and UV localized, or similar and UV
localized.  The profiles of the right-handed charged leptons
are always hierarchical and localized near the UV brane. We outline both cases below.
\begin{itemize}
\item (A) In the case where the left-handed and right-handed profiles are
hierarchical, they
 satisfy the following relations:
\begin{eqnarray}
f_L^i f_e^i \sim \frac{m_i^l}{\bar{Y} v_4}, \qquad
(O_{L,e})^{i,j}\sim\frac{f_{L,e}^i}{f_{L,e}^j},\qquad~ i<j.
\end{eqnarray}
where $f_{L,e}$ are profiles of the left-handed and right-handed
fields and $(O_{L,e})^{i,j}$ is  the intergenerational mixing. Then  the $a^l_{ij}$ become:
\begin{eqnarray}
a^l_{ij}\sim \frac{2}{3}\bar{Y}^2 (v_4^2 {R'}^2)
\sqrt{\frac{f_L^if_e^j}{f_L^jf_e^i}}.
\end{eqnarray}
This $a_{ij}^l$ are maximal  when $\displaystyle \frac{f_L^i}{f_L^j}\sim
\frac{f_e^i}{f_e^j}\sim \sqrt{\frac{m^l_i}{m_j^l}}$, i.e., when the
hierarchy of charged lepton masses gets equal contributions from the
 left-handed and right-handed fields.

\item (B) If right-handed profiles are hierarchical and left-handed
profiles are similar,  $f_L^1 \sim f_L^2 \sim f_L^3$,  the
profiles satisfy the following relations:
\begin{eqnarray}
f^i_L f_e^i &\sim& \frac{m^l_i}{\bar{Y} v_4}, \qquad
\frac{f_L^i}{f_L^j}\sim O(1) ,\qquad
 \frac{f_e^i}{f_e^j}\sim \frac{m_i^l}{m_j^l},\qquad~i<j,
\end{eqnarray}
and the the parameter $a^l_{ij}$ becomes:
\begin{eqnarray}
\label{hlfv2} a^l_{ij}\sim \frac{2}{3}\bar{Y}^2 (v_4^2 {R'}^2)
\sqrt{\frac{f_e^j}{f_e^i}}.
\end{eqnarray}
These flavor violating Higgs Yukawa couplings to leptons can also lead to
interesting collider signals for the decays of the fourth generation leptons, as
discussed in the next section.
\end{itemize}

\subsection{Tree Level $Z^0$ flavor violating couplings}

FCNC couplings of the $Z^0$ boson have been studied before in the
context of warped scenarios with 3 generations
\cite{Casagrande:2008hr}. These couplings arise basically from
two sources. First, the bulk profiles of the lowest-lying massive
gauge bosons (the SM $Z^0$ and $W^0$) are not flat, yielding non-trivial
and non-universal overlap integrals with the fermion profiles. Second,
even if the $Z^0$ and $W^0$ profiles were flat, there would still be a
non-universal correction to these couplings due to misalignments in
the fermion kinetic 
terms. In fact the correction has the exact same origin as the
misalignment $\hat{\Delta}^d_{2} $ in the Higgs sector shown in
Eq.~(\ref{threegenresult2}). 

For light quarks, the first source of misalignment dominates due to
Yukawa suppression of the fermion kinetic term misalignments. But for
heavier quarks, and specially fourth generation quarks, this last
source of flavor should dominate and this is the one we consider in
the following.

We can write the couplings of fermions with $Z^0$ as:
\bea
{\cal{L}}_{Z} =  \left [  g_L\
  \delta_{ij} + \left( \hat{\delta}_{g_L}\right)_{ij}\right] 
\bar{d}_L^i Z\!\!\!\!/\   d_L^j\ \  + \ \ \left[  g_R\
  \delta_{ij} + \left( \hat{\delta}_{g_R}\right)_{ij} \right]
\bar{d}_R^i Z\!\!\!\!/\  d_R^j\   +\  (d \leftrightarrow u),
\eea
where $\displaystyle g_L=\frac{g}{\cos{\theta}_W} (T_3-Q\sin{\theta_W}^2)$ and
$\displaystyle g_R=\frac{g}{\cos{\theta_W}} Q \sin{\theta_W}^2$ are the usual
diagonal SM couplings with $g$ the $SU(2)_L$ coupling constant, and $Q$
and $T_3$  the charge and the isospin of the quark in question. 
The corrections coming from the kinetic term misalignment are, for
the down quarks, 
\bea
\hat{\delta}^{kin}_{ g_L} &=& -\frac{gT^d_3}{\cos \theta_W}\ \ \hat{m}^{d\dagger} \hat{K}_{c_q} \hat{m}^d\  R'^2, \\  
\hat{\delta}^{kin}_{ g_R} &=&\ \  \frac{gT^d_3}{\cos \theta_W}\ \ \hat{m}^d \hat{K}_{c_d} \hat{m}^{d\dagger}\  R'^2.
\eea
where $\hat{m}^{d}$ is the fermion mass matrix before diagonalization,
${R'}^{-1}$ is the KK scale and $\hat{K}$ is a diagonal matrix whose
entries $K(c)$ were defined in Eq.~(\ref{Kc}). Upon diagonalization of
the fermion mass matrix in order to go to the physical basis, these
corrections will not be diagonal and will produce flavor violating
coupling for the $Z^0$ boson. The same mechanism 
applies in the up-sector.

Once in the physical basis, we can parametrize the off-diagonal
quark couplings in the Lagrangian by $\left (a_L^{u,d}\right )_{ij}$
and $\left ( a_R^{u,d}\right )_{ij}$, with   
 \begin{equation}
\label{aijLR}
{\cal{L}}_{ZFV} = -\frac{gT^d_3}{\cos \theta_W}  \left [  \left ( a_L^{d}\right )_{ij} 
\bar{d}_L^i Z\!\!\!\!/\   d_L^j - \left ( a_R^{d}\right )_{ij} 
\bar{d}_R^i Z\!\!\!\!/\  d_R^j  \right ] + (d \leftrightarrow u).
\end{equation} 
The $Z^0$ FCNC couplings $\left ( a_L^{u,d}\right )_{ij}$, $\left (
a_R^{u,d}\right )_{ij}$ can then be obtained from the same scan used
to obtain numerical values for the Higgs FCNC couplings. For example,
for the $(43)$ entries in the up and down sector, we find typical
ranges 
\bea
(a^u_L)_{43}={0.00350-0.0176},\ \ &&\ \  (a^u_R)_{43}={0.0274-0.0952}, \\
(a^d_L)_{43}={0.00356-0.0161},\ \  &&\ \ (a^d_R)_{43}={0.0209-0.0830}. 
\eea  
To obtain these values  we followed the same procedure explained
previously in the subsection {\it ``Numerical results for Higgs FCNC
  couplings''}.

\section{Phenomenology}
\label {sec:pheno}

\subsection{Bounds on Higgs-mediated FCNC couplings}

The off-diagonal Higgs Yukawa couplings induce FCNC, which 
affect many low energy observables and also give possible signatures
at colliders. In this section, we discuss first bounds  on Higgs
flavor violation coming from tree-level processes $\Delta F = 2$, such as
$K-\bar{K}$, $B- \bar{B} $, $D-\bar{D} $ mixing. We then study the
effects on loop processes, such as $b$ and $t$ flavor-changing decays,
as well as on   $Z\to b\bar{b},\ \tau^+\tau^-$. The radiative
processes are enhanced due to heavy quarks in the loop, and strong
off-diagonal Yukawa couplings.

\subsubsection{Tree-level processes}

The $\Delta F = 2$ process can be described by the general
Hamiltonian \cite{UTfit,BurasWeakHamiltonian}
\begin{eqnarray}
{\cal H}_{eff}^{\Delta F =2} = \sum_{a=1}^{5} C_a Q_a^{q_i q_j} +
\sum_{a=1}^3 \tilde{C}_a \tilde{Q}_a^{q_i q_j}, 
\end{eqnarray}
with
\begin{eqnarray}
Q_1^{q_i q_j} &=& \bar{q}^\alpha_{jL}\gamma_\mu
q_{iL}^\alpha\bar{q}^\beta_{jL}\gamma^\mu q^\beta_{iL}, \quad
Q_2^{q_i q_j} = \bar{q}^\alpha_{jR} q_{iL}^\alpha
\bar{q}^\beta_{jR} q_{iL}^\beta, \quad Q_3^{q_i q_j} =
\bar{q}^\alpha_{jR}q_{iL}^\beta \bar{q}_{jR}^\beta q_{iL}^\alpha
, \\ \nonumber Q_4^{q_i q_j} &=& \bar{q}^\alpha_{jR}q_{iL}^\alpha
\bar{q}_{jL}^\beta q_{iR}^\beta, 
\quad 
Q_5^{q_i q_j}=
\bar{q}^\alpha_{jR} q_{iL}^\beta \bar{q}^\beta_{jL} q_{iR}^\alpha ,
\end{eqnarray}
where $\alpha, \beta$ are color indices. The operators $\tilde{Q}_a$
are obtained from $Q_a$ by exchange $L \leftrightarrow R$. For
$K-\bar{K}$ , $B_d-\bar{B}_d $, $B_s-\bar{B}_s$, $D-\bar{D}$
mixing, $q_i q_j = s d$, $b d$, $ b s$ and $ u c$ respectively.

Exchange of the flavor-violating Higgs bosons gives rise to new contribution to
$C_2$,
$\tilde{C}_2$ and $C_4$ operators \cite{BlankedeltaF2}.  These contributions have been
included in
\cite{toharia1}, and the basic bounds on the coefficients are not
altered. We present them here, for completeness, in a more general
fashion, with no relation to the possible numerical values of the
entries in the Higgs Yukawa mass matrix. We use the model-independent
bounds on BSM contributions as in \cite{UTfit}, and present coupled
constraints on the Higgs flavor violating Yukawa couplings
parametrized by the $a_{ij}$ couplings and  the Higgs mass $m_h$.  

 \begin{itemize}

\item {$K^0-\bar{K}^0$ mixing}

The coefficients $C_2$,   $\tilde{C}_2$ and $C_4$ will set limits on
the real and imaginary of the Yukawa couplings  $a^d_{12}, \ a^d_{21}$,
and their product. Specifically, for the values of parameters used in
the previous sections, we obtain, from $\Delta M_K$, respectively: 
\begin{eqnarray} 
&&|(a_{12}^d)|\left(\frac{400~ {\rm GeV}}{m_h}\right)\le  0.62, \quad
  |(a_{21}^d)|\left(\frac{400~ {\rm GeV}}{m_h}\right) \le  0.62,
  \\ \nonumber   
&&|(a_{12}^d a_{21}^d)|\left(\frac{400~ {\rm GeV}}{m_h}\right)^2\le  (0.35)^2.
\end{eqnarray}

The bounds obtained from $\epsilon_K$ are very stringent, and restrict
the phases of the off-diagonal Higgs Yukawa couplings: 
\begin{eqnarray}
&&\text{Im}(a_{12}^d)^2  \left(\frac{400~ {\rm GeV}}{m_h}\right)^2 \le  \left
(4.6 \times 10^{-2}\right)^2, \quad
\text{Im}(a_{21}^d)^2\left(\frac {400~ {\rm GeV}}{m_h}\right)^2 \le  \left(4.6
\times 10^{-2}\right)^2\nonumber\\
&&\text{Im}(a_{12}^d a_{21}^d)\left(\frac{400~ {\rm GeV}}{m_h}\right)^2 \le 
\left(2.2 \times 10^{-2}\right)^2
\end{eqnarray}

\item {$D^0-\bar{D}^0$ mixing}

The $D^0-\bar{D}^0$ mixing constrains the ${(12,\ 21)}$  off-diagonal entries
in the up-quark flavor changing mixings.
\begin{eqnarray}
&&|(a_{12}^u)|\left(\frac{400~ {\rm GeV}}{m_h}\right) \le  0.71, \quad
|(a_{21}^u)|\left(\frac{400~ {\rm GeV}}{m_h}\right) \le  0.71, \nonumber\\
&&|(a_{12}^u a_{21}^u)|\left(\frac{400~ {\rm GeV}}{m_h}\right)^2 \le  \left
(0.47 \right) ^2.
\end{eqnarray}

\item {$B^0_d-\bar{B}^0_d$ mixing}

The mass mixing in the $B^0_d-\bar{B}^0_d$ is fairly constrained,
resulting in bounds on the ${(13, \ 31)}$ entries in the down-quark flavor
changing mixings. 
\begin{eqnarray}
&&|(a_{13}^d)|\left(\frac{400~ {\rm GeV}}{m_h}\right) \le  0.54, \quad 
|(a_{31}^d)|\left(\frac{400~ {\rm GeV}}{m_h}\right) \le  0.54, \nonumber \\
&&|(a_{13}^d a_{31}^d)|\left(\frac{400~ {\rm GeV}}{m_h}\right)^2 \le 
\left(0.35\right)^2. 
\end{eqnarray}

\item {$B^0_s-\bar{B}^0_s$ mixing}

The mass mixing in the $B^0_s-\bar{B}^0_s$ is less restricted than in the
$B_d^0$ sector, resulting in bounds on the ${(23, \ 32)}$ entries in the
down-quark flavor changing mixings. At first, these bounds may not
appear useful; however, one must note that the matrix entries $a_{ij}$
are not otherwise constrained ({\it e.g.}, by unitarity). 
\begin{eqnarray}
&&|(a_{23}^d)|\left(\frac{400~ {\rm GeV}}{m_h}\right) \le  1.1, \quad
|(a_{32}^d)|\left(\frac{400~ {\rm GeV}}{m_h}\right) \le  1.1, \nonumber\\
&&|(a_{23}^d a_{32}^d)|\left(\frac{400~ {\rm GeV}}{m_h}\right)^2 \le 
\left(0.64\right)^2.
\end{eqnarray}

\end{itemize}
With the exception of $\epsilon_K$, these bounds are not too
restrictive over the estimated size of the flavor violating couplings
of the Higgs as our numerical evaluation show, even for lighter $m_h \simeq 
200$ GeV.

In what follows, we compare the tree-level bounds with precision
bounds coming from loop-generated processes including a heavy fermion
in the loop. 
\subsubsection{One-loop processes}

We evaluate flavor-violating radiative type processes of the form $q_i
\to q_j \gamma$, and $l_i \to l_j \gamma$  as well as $Z \to b {\bar
  b}$ and $Z \to \tau^+ \tau^-$. Though occurring at one-loop level,
these processes are tightly constrained experimentally.
For a recent calculation of these warped penguin diagrams due to
radiative exchanges of heavy KK states see \cite{Csaki:2010aj}.
In our scenario each process receives additionally non-universal
contributions from the fourth generation quarks or leptons and Higgs
bosons running in the loop. 

The contribution is enhanced for couplings with the third generation, as
the FCNC couplings are larger. The basic process is illustrated in
Fig. 3, where $F$  represent fourth generation quarks or leptons,
$f_i,\,f_j$, second or third generation  quarks or leptons, and $h$ is
the Higgs boson . For instance,  for $b \to s \gamma$, $F=b^\prime$,
$f_i=b$ and $f_j=s$ quarks, while for $Z \to \tau^+ \tau^-$,
$F=\tau^\prime$, and $f_i=\tau^+,\, f_j=\tau^-$. We analyze each
process in detail.

\begin{figure}[t]
\center
\begin{center}
$
	\begin{array}{ccc}
\hspace*{-0.4cm}
	\includegraphics[width=1.8in,height=1.2in]{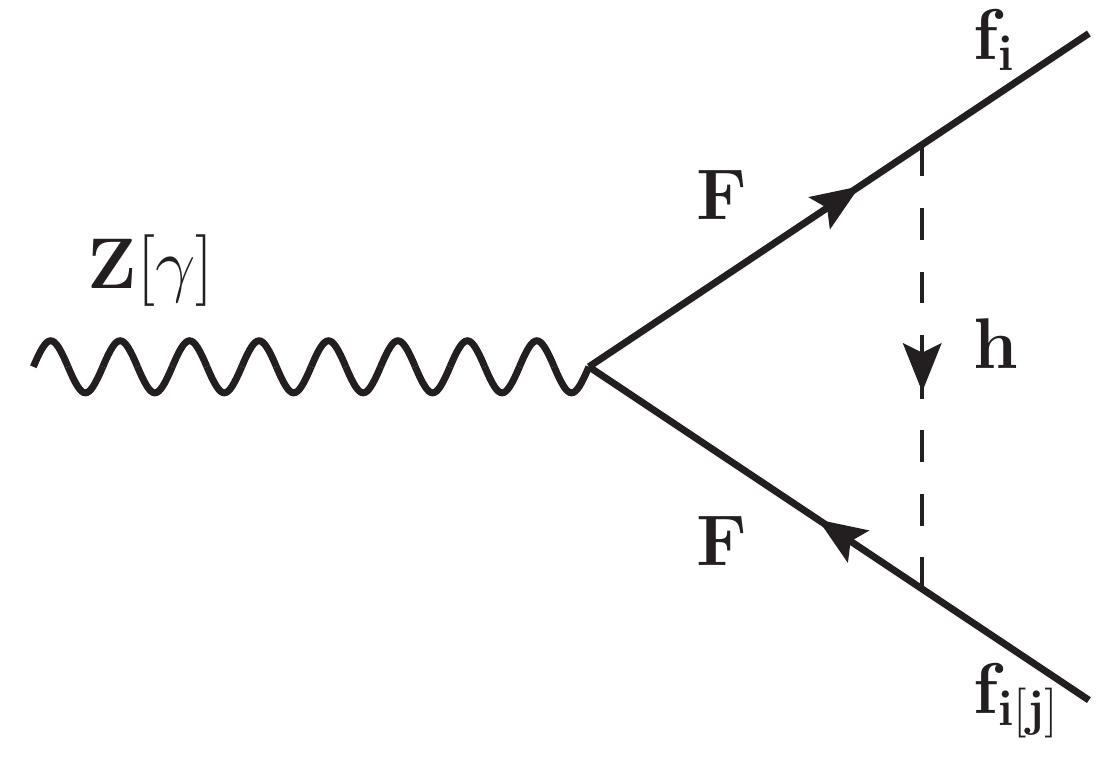}
&\hspace*{0.2cm}
	\includegraphics[width=1.8in,height=1.2in]{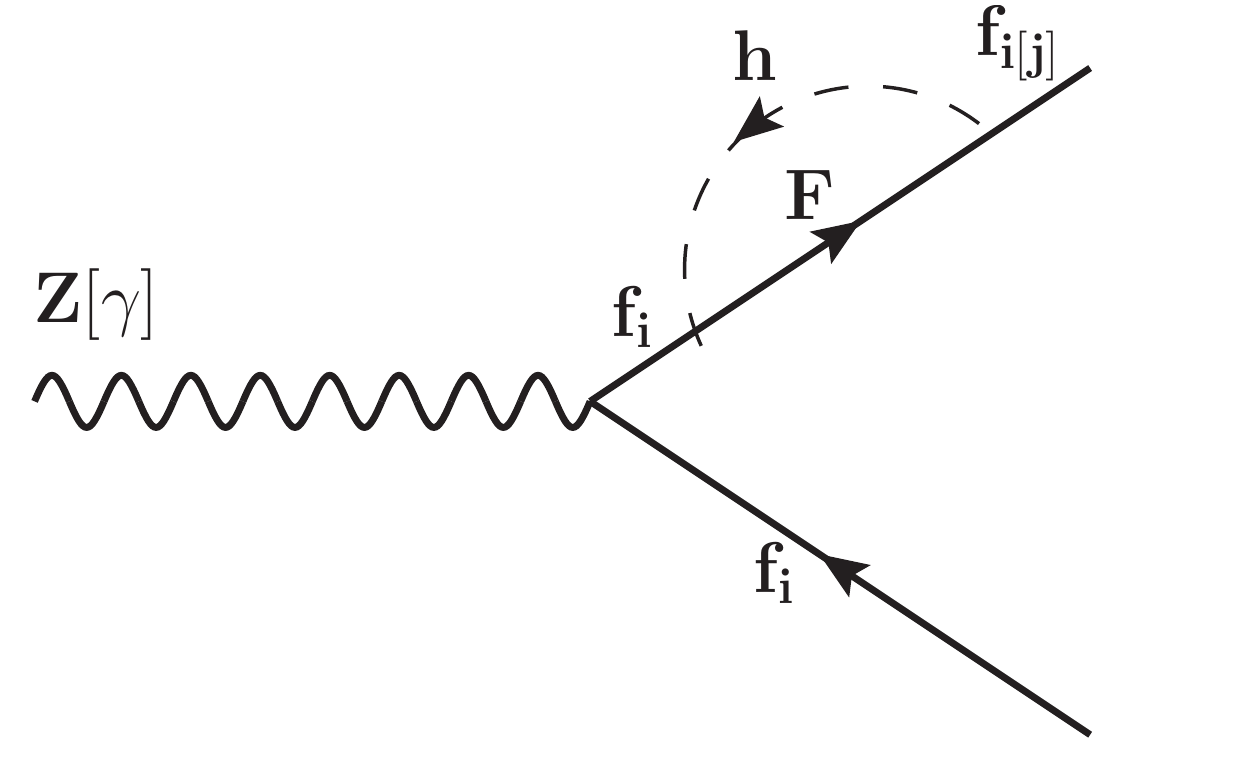}
&\hspace*{0.2cm}
\includegraphics[width=1.8in,height=1.2in]{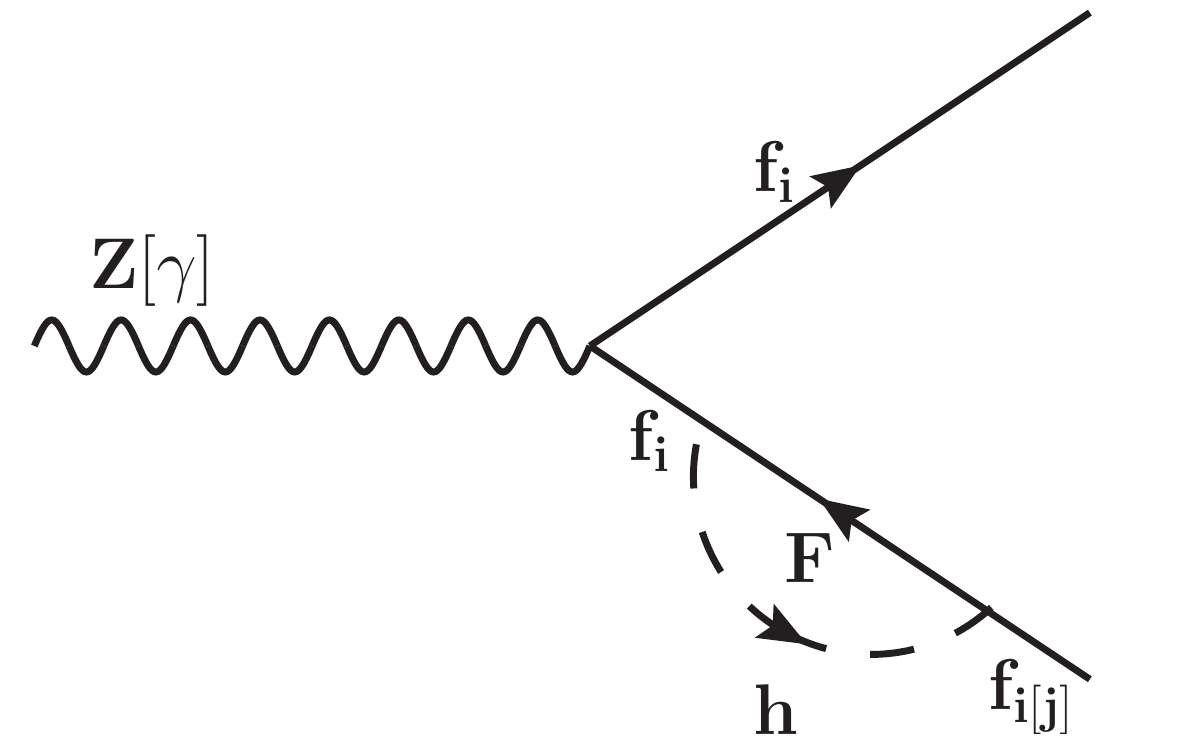}
        \end{array}$
        \end{center}
\caption{Generic loop diagrams enhanced by FCNC couplings
          between Higgs boson and 4th generation fermions. Here $F$
          stand for a 4th generation quark (or lepton), while $f_i,
          f_j$ are 2nd or 3rd generation quarks (or leptons). The
          left-hand side graph is the vertex diagram, while the other
          two are self-energy diagrams. } 
\label{gff}
\end{figure}

\begin{itemize}
 
 \item{$b\to s \gamma $ induced by Higgs FCNC couplings.}

The decay rate of $b \rightarrow s\gamma$ is
\begin{equation}
\Gamma(b\rightarrow s
\gamma)=\frac{\langle\mathcal{M}^{2}\rangle}{16\pi
m_{b}^{3}}\sqrt{m_{b}^{4}+m_{s }^{4}-2m_{b}^{2}m_{s}^{2}}.
\end{equation}
where the most dominant term in the matrix element 
 $\mathcal{M}^{2}$ is 
\begin{equation}
 \langle\mathcal{M}^{2}\rangle=\frac{em_{b'}^{4}m_{b}m_{s}}{(24\pi^{2}v_{4}^{2}
)^{2}
}|a_{42}^{d}a_{34}^{d}|^{2}(m_{b}^{4}+m_{s}^{4}-2m_{b}^{2}m_{s}^{2})C_{0}^{2}
\Big(P_{1}^{2},P_{2}^{2},(P_{1}+P_{2})^{2},m_{b'}^{2},m_{h}^{2},m_{b'}^{2}
\Big),
\end{equation}
and where $C_0$ is a three point integral as defined in Looptools \cite{Hahn:2010zi} (see
Appendix A for a more detailed calculation).

Using the experimental value of the branching ratio of $\bar{B}\rightarrow
X_{s}\gamma$
\begin{equation}
 Br(\bar{B}\rightarrow X_{s}\gamma)=(3.55\pm0.24\pm0.09)\times10^{-4}, 
\end{equation}
we can put a bound on $a_{ij}$'s such that $|a_{42}^{d}a_{34}^{d}|\leq
1.3$. This is a very conservative bound. If we require the branching ratio to be
the sum of the SM and the new physics contribution, and use the NLO
result Br$(\bar{B}\rightarrow
X_{s}\gamma)_{E_\gamma>1.6~GeV}=(3.60\pm0.30)\times10^{-4}$
\cite{BarShalom:2011zj}, we obtain $|a_{42}^{d}a_{34}^{d}|\leq 0.45$.  
 These values start to be quite restrictive, as compared to the
 expected size predicted by our scenario
 $|a_{42}^{d}a_{34}^{d}|\simeq 0.85$ (obtained from our numerical scan).

\item{$ \tau \to \mu \gamma, \ \tau \to e \gamma, \ \mu \to e \gamma$
  induced by Higgs FCNC couplings} 

The same operators will contribute to lepton FCNC decays. The experimental
limits on these processes are \cite{particledata}
\begin{eqnarray}
Br(\tau \to \mu \gamma)&\leq &4.4 \times 10^{-8}, \nonumber\\
Br(\tau \to e \gamma)&\leq &3.3 \times 10^{-8},\nonumber \\
Br(\mu \to e \gamma)&\leq &1.2 \times 10^{-11}.
\end{eqnarray}
The Higgs mediated diagrams with a heavy $\tau^\prime$ in the loop will yield
limits on the $a^l_{ij}$ parameters. Specifically, we get
\begin{eqnarray}
|a_{34}^l a_{42}^l |&& \leq 0.11, \quad |a_{34}^l a_{41}^l |\leq 1.45, \quad 
|a_{24}^l a_{41}^l |\leq 0.002. 
\end{eqnarray}
We also calculated the $a^{l}_{ij}$ values by using the two different
scenarios. In scenario (A) where both the left-handed and right-handed
profiles are hierarchical, we have 
\begin{eqnarray}
|a_{34}^l a_{42}^l|=|a_{34}^l a_{41}^l |=|a_{24}^l a_{41}^l|\simeq 0.0065.
\end{eqnarray}
However, in scenario (B) where right-handed profiles are hierarchical and
left-handed profiles are not,  we get
\begin{eqnarray}
|a_{34}^l a_{42}^l|\simeq 0.0016, \quad 
|a_{34}^l a_{41}^l |\simeq 0.00011, \quad
|a_{24}^l a_{41}^l|\simeq 0.00045.
\end{eqnarray}

Using the $a_{ij}^{l}$ values in scenario (A) and ${\bar Y} =3$ we calculated the branching
ratios as Br$(\tau \to \mu \gamma)= 1.4\times 10^{-10}$,~Br$(\tau \to e
\gamma)=6.7 \times 10^{-13}$ and 
Br$(\mu \to e \gamma)=6.2 \times 10^{-11}$.

For scenario (B) (keeping ${\bar Y} =3$) we have Br$(\tau \to \mu \gamma)=7.8
\times 10^{-12}$,~Br$(\tau \to e \gamma)=1.9 \times 10^{-16}$ and 
Br$(\mu \to e \gamma)=2.9 \times 10^{-13}$. The predicted size of
flavor ciolating $\tau$ decays  lies just below experimental bounds,
but the branching ratio for $\mu \to e \gamma$ is above the
experimental bounds in scenario (A), and therefore sets some bounds or
pressure on our scenario. More stringent limits can be set when
(expected) new experimental results become available.

\item{$ t \to c \gamma$ induced by Higgs FCNC couplings}

Using the formalism from $b \to s\gamma$ we can estimate the branching ratio for
$t \to c \gamma$. We obtain
\begin{equation}
Br( t \to c \gamma)=1.55 \times 10^{-9}\Big
[|a^u_{42}a^u_{34}|^2+|a^u_{43}a^u_{24}|^2+0.25\Re \left (a^u_{24}a^u_{43}a^u_{42} a^u_{34} \right) \Big ].
\end{equation}
which for our values of the scanned Higgs couplings becomes Br$( t \to c
\gamma)=1.33 \times 10^{-12}$, too small to be detected anytime soon,
and comparable to the SM estimate Br$( t \to c \gamma)=4.5 \times 10^{-13}$
\cite{Eilam:1990zc}.

\item{$Z\to b\bar{b}$ decay and $Z \to \tau^+\tau^-$}

For completeness we also computed the loop corrections to $Z\to
b\bar{b}$ decay and $Z \to \tau^+\tau^-$. The $b'$ and $\tau'$ running
in these loops make these diagrams larger than the corresponding case
with three generations but are still too small to place any useful
bound on the Higgs FCNC couplings. (See Appendix A for details.)

\end{itemize}

\subsubsection{Higgs production and decay}
\label{Hprod-decay}

\begin{figure}[t]
\center
\vspace{-1cm}
\includegraphics[width=15cm,height=8.5cm]{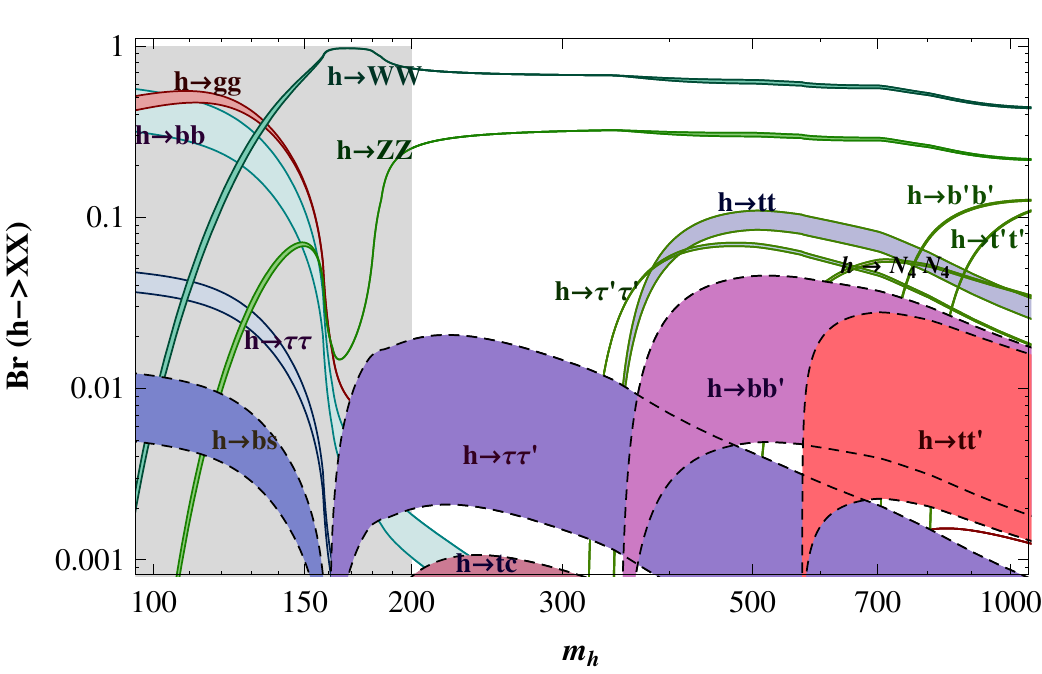}
\vspace{-.2cm}
\caption{Decay branching fractions of the Higgs scalar in a warped scenario
  with four generations of fermions. The bands represent 50\%
  likelihood for the branching ratio, according to our numerical scan,
  as explained in the text.  The gray region below $200$ GeV is excluded by both
  Tevatron and LHC. The flavor anarchy setup (masses and mixings
  explained through fermion localization, with random 5D Yukawa
  couplings) predicts generic FV couplings of the Higgs, leading to a
  few new interesting decay channels such as $h\to bb'$ and $h \to
  \tau\tau'$. The masses chosen for this plot are $m_{b'}=350$ GeV,
  $m_{t'}=400$ GeV, $m_{\tau'}=160$ GeV and $m_{N_4}=250$ GeV ($N_4
  \equiv \nu_{\tau}'$), and the KK scale is $(R')^{-1}= 1500$ GeV. }     
\label{BRplot}
\end{figure}

The Higgs emerging in RS with 4 generations is in fact quite similar
to the SM Higgs with 4 generations (SM4). The tree level couplings are still
proportional to the masses of the particles it couples to. One of the 
main differences between four generations and three generations, from the
Higgs perspective, is the new radiative contributions to the coupling of Higgs
to photons and gluons. This last coupling is typically enhanced by a
factor of $\sim {\cal O}(3)$ (due to three heavy quarks running in the
loops instead of only the top quark), and since the Higgs is mainly produced
through gluon fusion at hadron colliders, one expects roughly an
enhancement in production cross section of $\sim{\cal O}(9)$. Of
course this enhancement must be carefully calculated as it is still
sensitive to the relative mass between the Higgs and the heavy
quarks. In any case the production cross section of this Higgs with four
generations will allow the appearance of many more Higgs bosons than
predicted by the minimal SM. Therefore the SM Higgs bounds from Tevatron
now become quite stringent, and even early LHC data allows exclusions in
the parameter space \cite{Gunion:2011ww, Higgs mass limits}. In particular it
seems
that a Higgs mass smaller than $200$ GeV is already excluded
by hadron collider bounds (assuming that no new decay channels exist
for the Higgs). We will take $200$ GeV as a lower bound for
our Higgs scalar and study the possible decay channels that such a
heavy Higgs could have. The bands represent 50\% likelihood for the branching ratio, as given in our numerical scan.  (That is, 25\% of all the parameter points from the numerical scan lie below and 25\% lie above the shown interval.) The results are shown in Figure \ref{BRplot},
where the branching fraction for each channel is presented. Not
surprisingly the dominant decay modes for such a heavy Higgs
($m_h>200$ GeV) are the usual decay channels, namely $h\to W^\pm W^\mp$ and $h\to
Z^0Z^0$ where both $W$ pairs and $Z^0$ pairs are on-shell. These are the
same dominant channels as in the SM; of course once above threshold
the Higgs should also decay into pairs of heavy fermions. The typical
expectation for models with four generations is that Higgs decays into
$t\bar{t}$, $t'\bar{t'}$, $b'\bar{b'}$, ${\tau'}^+ {\tau'}^-$ (fourth
generation charged lepton pair) or $\nu_{\tau}' \nu_{\tau}'$ (fourth generation
neutrino pair) will all have branchings similar to the branching of
$h\to t\bar{t}$, given that the masses of these fermions should
typically be in the hundreds of GeV (except maybe the $\nu_{\tau}'$).   
That yields  branching fractions at the $10\%$ level, and this is
confirmed in Figure \ref{BRplot}. 

The new and very interesting result is the prediction of sizable
branching fractions for exotic decays of the Higgs into 
fermion pairs of different flavor. In particular we observe that $h\to \tau
\tau'$,
$h\to bb'$ and $h\to tt'$ are among the most important new flavor
violating channels, a fact not surprising since for heavier fermions
one expects larger couplings to the Higgs. An interesting remark for
these new channels is that the threshold mass at which they become
kinematically allowed is basically set by the mass of the heaviest
fermion. This means that while some or most of the flavor diagonal
decays into fermions might be closed, there are good chances of
an open channel such as $h\to\tau\tau'$ or $h\to bb'$.
For the chosen parameters (KK scale of $1/R'=1500$ GeV and typical 5D
Yukawa couplings of ${\cal O}(2)$) we obtain generic flavor violating
Higgs couplings which place the branching ratios of these exotic decay modes
on the order of $10^{-2}$. Note that since the flavor
violating couplings scale as $(\bar{Y} R')^2$, the branching ratios should
in turn scale as $(\bar{Y} R')^4$, showing great sensitivity to both
the 5D Yukawa couplings and the KK scale. 

The production cross section at the LHC of a heavy Higgs of ~$200-400$ GeV,
in a scenario with fourth generation quarks is expected to be about
~$50-70$ pb \cite{particledata}. Since the new exotic decays have branching ratios at
the
percent level, one expects the cross section of these modes to be
somewhere near $500$ fb. This means that with 1 or 2 fb$^{-1}$
of integrated luminosity at the LHC (early stages) one could have at
least a few hundred of these events.  
Of course given the large production cross section, there would be no
problem in quickly discovering the Higgs via the four lepton mode
($h\to Z^0Z^0\to 4l$) or maybe through ($h\to W^\pm W^\mp $). With the Higgs mass
properly set, a complementary search for some of the new exotic
channels should be much easier. 

Of particular interest is the mode $h\to \tau\tau'$ since it may
actually compete as the main production mechanism for the fourth generation
charged lepton. If $m_h<2m_{\tau'}$, the decay into pairs of $\tau'$
is forbidden and so the other possible production for heavy leptons is
through s-channel processes involving electroweak bosons
\cite{Carpenter:2010bs} and their KK partners
\cite{Burdman:2009ih}. The typical cross section for $\tau'- \nu_{\tau}'$
production via $s$-channel $W$ is ~$10-100$ fb \cite{Carpenter:2010bs},
which means that the flavor violating production through $s$-channel
on-shell Higgs of $\tau^\pm {\tau'}^\mp$ can be a few times larger than this. 
The subsequent decay of the ${\tau'}^\mp \to \nu_{\tau}' W^\mp$, and then of $\nu_{\tau}' \to W
l$ should give a signal of $pp\to h\to \tau^\pm {\tau'}^\mp\to \tau^\pm W^\mp W l$,
where all particles are produced and decayed on-shell. The signs of the
second $W$ and the charged lepton $l$ is not fixed and depends on the
nature of $ \nu_{\tau}'$. One would look for same sign dilepton events coming from 
leptonic decays of the first $W$ along with the last lepton of the
chain. This type of signature is quite clean thanks to the minimal
background and would in principle allow for easy confirmation of the
signal, which could become the discovery signal for the $\tau'$ along
with the confirmation of Higgs flavor violating couplings.

Another interesting decay mode, if kinematically allowed is $h\to
bb'$, where the $b'$ would subsequently decay as $b'\to q\,W$ or $b'\to
b\, Z^0 $. In the first possibility, $q$ stands for $t$ if kinematically
allowed, and for $c$ or $u$. The partial width of these channels
depend on the size of the CKM4 angles $V_{tb'}$, $V_{cb'}$
and $V_{ub'}$ which are typically constrained to be small 
\cite{Alok:2010zj}. A channel which could compete is $b'\to b\,Z^0 $, since in
the
RS scenario under study these flavor violating couplings appear at
tree-level, in a similar fashion as in the Higgs sector
\cite{agashe,Casagrande:2008hr}. Thus depending on the decay
branching ratios of the $b'$ heavy quark (see next section) the events could be
$pp\to h\to bb' \to b W^- t \to bbW^\pm W ^\mp\ \ $ or $\ \ pp\to h\to bb' \to b W^-
j \ \ $ or $\ \ pp\to h\to bb' \to b b Z^0 $.
A careful study of these signals and their background is beyond the
scope of this work, but we should  mention that a clear prediction
of our scenario is that the $h-b-b'$ coupling is highly asymmetric
(see Eq.~(\ref{adest})) with a definite preference for $h\to 
b'_R b_L$ decay over the $h\to b'_L b_R$. Thus one should also look
for the angular correlations in the signals in order to search for
this asymmetric property of the couplings (see refs. \cite{Krohn:2011tw} for
studies along these lines).

\subsubsection{Heavy fermion decays}

\bit
\item Heavy quark decays
\eit

If the Higgs masses are lighter than the masses of the fourth generation
fermions, channels in which the heavy fermions decay to the Higgs boson and a fermion
from one of the lighter families are open. Pair production of heavy quark flavors is
expected to have a cross section of $ \sim 4-4.5$ pb for a mass of $500$ GeV{\footnote{ The
    cross sections are estimated based on QCD effects only, and are
    based on approximate knowledge of PDF, thus should be only seen as
    indicative.}} at the LHC with $\sqrt{s}=14$ TeV
\cite{Cacciari:2008zb},  thus should be within reach, and their
properties would then become apparent.    As the FCNC couplings of the
Higgs to 
the fermions are proportional to fermion masses, the dominant decays would be to
the third generation fermions. The flavor violating couplings of Higgs will lead
to tree-level decays
$t'\rightarrow th$ and $b'\rightarrow bh$ in the kinematically allowed regions
 $m_{t'}>m_{h}+m_{t}$ and $m_{b'}>m_{h}+m_{b}$. The decay rates
for these processes are calculated as
\begin{eqnarray}
\Gamma(Q_{j}\rightarrow q_{i}h)&=& \frac{1}{16\pi m_{j}^{3}}\sqrt{
m_{i}^{4}+m_{j}^{4}+m_{h}^{4}-2m_{i}^{2}m_{j}^{2}-2m_{i}^{2}m_{h}^{2}-2m_{j}^{2}
m_{h}^{2}}\nonumber\\
&\times&\Big [(\mid a_{ij}^{u(d)}\mid^2+\mid
a_{ji}^{u(d)}\mid^2)(m_{j}^{2}+m_{i}^{2}-m_{h}^{2})+4 \Re(a_{ij}^{u(d)}a_{ji}^{
u(d) } )m_{i}m_{j} \Big ]\frac{m_{i}m_{j}}{v_{4}^{2}}.\ \ \ \ \label{Qqh}
\end{eqnarray}
 These decays can have significant decay width, and branching ratios. 
By comparison, the other dominant two body decay modes are $t^\prime\rightarrow
bW$ and $b^\prime \rightarrow tW$, given by  \cite{Atwood:2011kc}
\begin{eqnarray}
 \Gamma(Q_{j}\rightarrow q_{i}W)&=& \frac{\alpha\left|
V_{ji}\right|^2}{16 M_{W}^{2}m_{j}^{3}} \sqrt{
m_{i}^{4}+m_{j}^{4}+M_{W}^{4}-2m_{i}^{2}m_{j}^{2}-2m_{i}^{2}M_{W}^{2}-2m_{j}^{2}
M_{W}^{2}}\nonumber\\
&\times&\bigg(
m_{i}^{4}+m_{j}^{4}-2M_{W}^{4}-2m_{i}^{2}
m_{j}^{2}+m_{j}^{2}
M_{W}^{2}, \label{Qqw}
\bigg),
\end{eqnarray}
by substituting the corresponding quarks in the two body decays. 
The flavor-changing couplings of quarks to the $Z^0$ boson  allow FCNC quark decays via the process $Q \to q Z^0$. The branching ratio is  \cite{Casagrande:2008hr}
\begin{eqnarray}
 \Gamma (Q_j \to q_i Z^0) 
   &=& \frac{\alpha T_3^2}{8 M_{Z}^{2}\cos^2 \theta_W m_{j}^{3}} \bigg({\sqrt{
m_{i}^{4}+m_{j}^{4}+M_{Z}^{4}-2m_{i}^{2}m_{j}^{2}-2m_{i}^{2}M_{Z}^{2}-2m_{j}^{2}
M_{Z}^{2}} }\bigg) \nonumber\\
&\times& \left \{
 \left (m_{j}^{2}-m_{i}^{2} \right)^2+M_{Z}^{2}\left (m_{j}^{2}-2M_Z^2\right )         
  \bigg [\left| \left( a_L^{u,d} \right)_{34} \right|^2
    + \left| \left( a_R^{u,d} \right)_{34} \right|^2 \bigg] \right. \nonumber\\
   & +&\left. 12\, m_i m_j^3\,
    {\Re}\big[ \left( a_L^{u,d} \right)_{34}^*
     \left( a_R^{u,d} \right)_{34} \big] \right\}, \label{Qqz}
     \end{eqnarray}
with $T_3$ the third quark isospin component and with the flavor-changing
couplings $a_L^{u,d}$ and $a_R^{u,d}$ as defined in 
given as in Eq.~(\ref{aijLR}).
We define the total width to be the sum of the dominant two body-decays
\begin{equation}
 \Gamma(Q_{j}\rightarrow 2X)=\Gamma(Q_{j}\rightarrow
q_{i}W)+\Gamma(Q_{j}\rightarrow q_{i}h)+\Gamma(Q_{j}\rightarrow q_{i}Z^0).
\end{equation}
Although the decays $Q_j \to q'_i W$, $Q_j \to q_i Z$ and $Q_j \to q_i
h$, $i=1,2$ should be subdominant due to CKM and Yukawa suppression, for
completeness we include them in our numerical calculations and plots.

\begin{figure}[t]
\center$
	\begin{array}{ccc}
\hspace*{-0.4cm}
	\includegraphics[width=2.2in,height=3.2in]{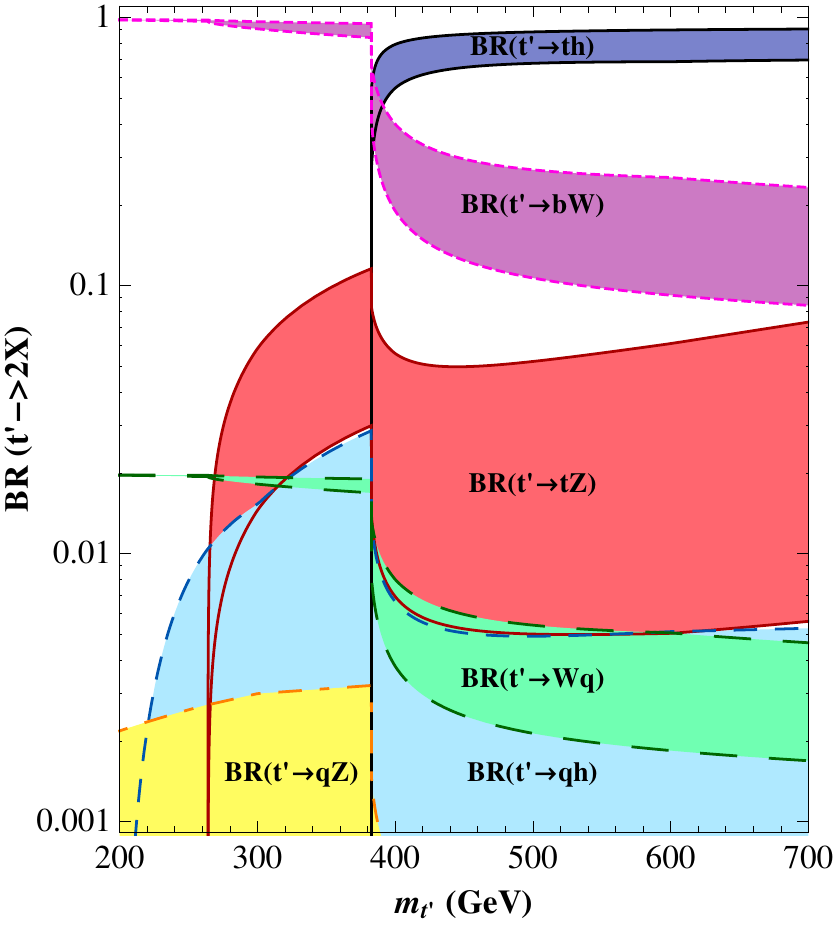}
&\hspace*{-0.2cm}
	\includegraphics[width=2.2in,height=3.2in]{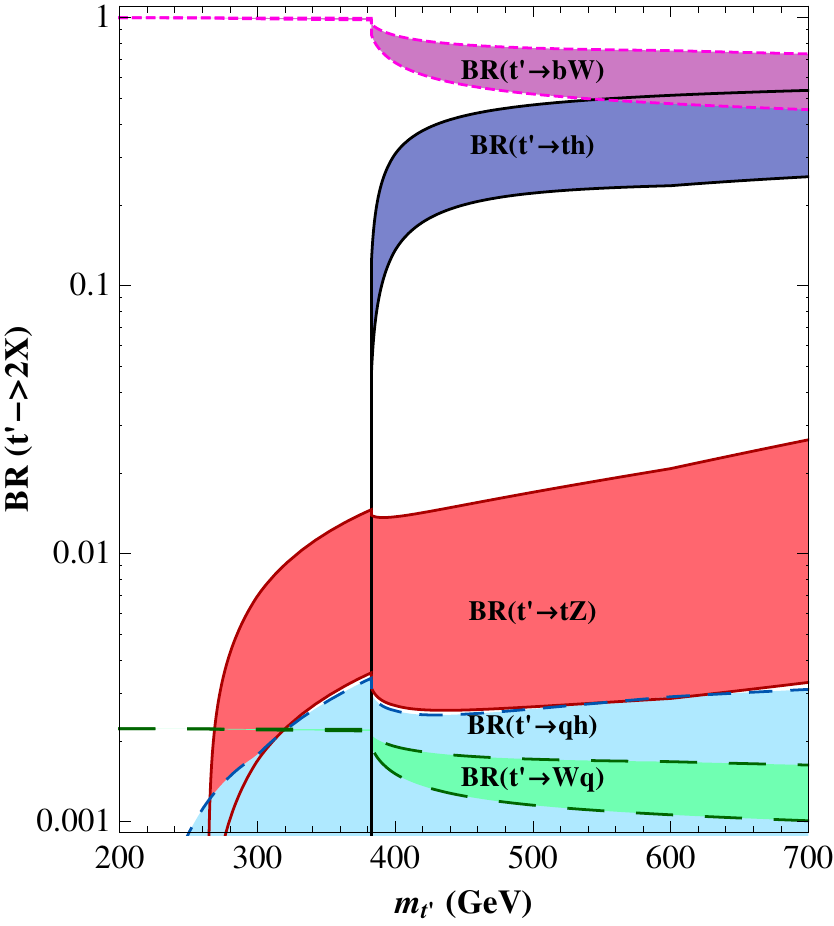}
&\hspace*{-0.2cm}
        \includegraphics[width=2.2in,height=3.2in]{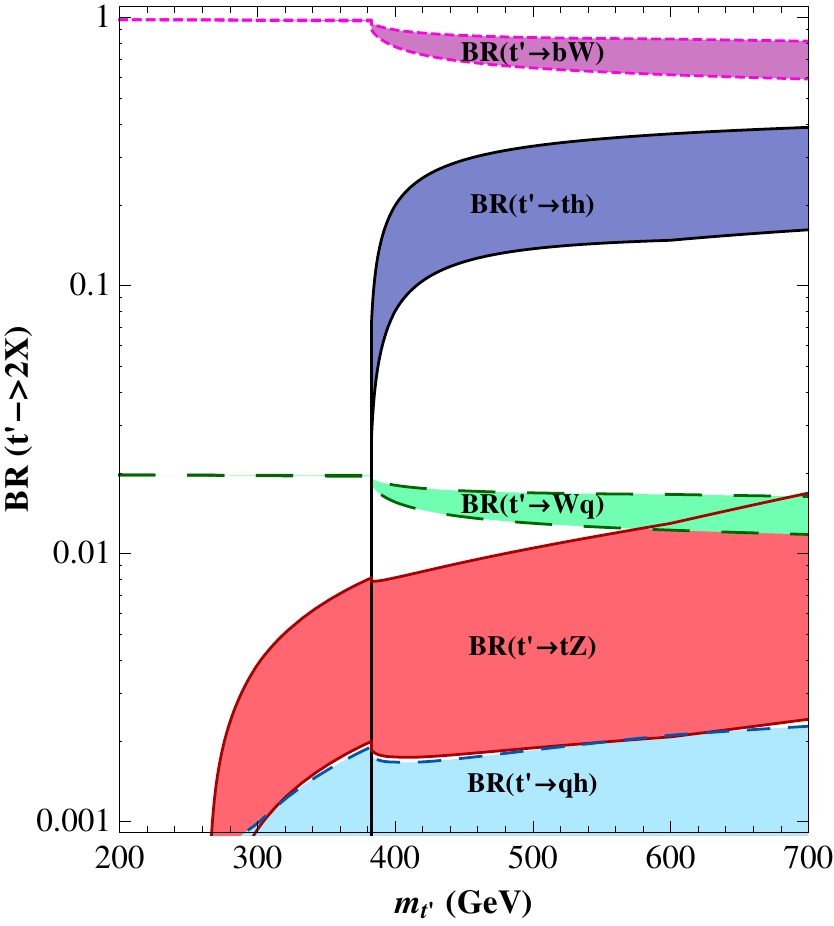}\\
        \end{array}$  
\caption{Branching ratios for 2-body $t^\prime$ decays with CKM4 mixing
  angle $V_{t^\prime b}=0.1$ and KK scale $\ \ R'^{-1}(\equiv M_{KK})=1.5$ TeV (left panel),
  $V_{t^\prime b}=0.3,\ R'^{-1}=1.5$ TeV (middle panel) and
  $V_{t^\prime b}=0.1,\ R'^{-1}=3$ TeV (right panel). We take
  $V_{t^\prime s}=V_{t^\prime d} = 0.01$ and $m_h= 200$ GeV
  throughout. The bands represent 30\% likelihood for the branching
  ratio, according to our numerical scan, as explained in the text.}   
\label{BRtprim}
\end{figure}

In Figure \ref{BRtprim} we illustrate the branching ratios for the
$t^\prime$ quark for $m_h=200$ GeV (still allowed by present bounds on
the Higgs in the presence of four  generations \cite{Higgs mass limits}) for two
choices of KK mass scales, $R'^{-1}=1.5$ TeV and $R'^{-1}=3$ TeV, and for two
choices of the CKM4 mixing involved here, i.e $V_{t^\prime b}=0.1$ and
$V_{t^\prime b}=0.3$. The latter will affect the tree-level decay $t^\prime
\to b\, W$, typically assumed to be the dominant decay for the usual 
choice $m_{t^\prime}-m_{b^\prime}\lsim 50$ GeV. The characteristic
bands appearing in these figures are due to the fact that the flavor
violating couplings for both Higgs and $Z^0$ are obtained from numerical
scans, performed for different values of the heavy quark masses. To
visualize the generic region in parameter space that the branchings
should cover, we show the interval of couplings inside which
30$\%$ of all the generated points lie,  such that 35$\%$ lie below
that interval and 35$\%$ lie above. This procedure will define
``bands'' in the figures which should be understood 
as the generic region predicted by flavor anarchy, containing 30$\%$
of the random points (with 35$\%$ of the points lying above the band
and 35$\%$ lying below).

We compare  the
dominant branching ratios for tree level decays: $t^\prime \to b\, W$,
$t^\prime \to t\, h$ and $t^\prime \to t\,Z^0$, and also  the
subdominant decays $t^\prime \to q'\,W$, $t^\prime \to q\,Z$ and
$t^\prime \to q\,h$ with $q'=d,s$ and $q=c,u$. 
Compared to these tree-level decays, the branching of
loop-induced processes such as Br$(t^\prime \to t \,\gamma) \simeq 
{\cal O}(10^{-7})$ are much smaller. 
In all three plots we observe the importance of the decay rate $t'\to
t\,h$, which will generically dominate for a KK scale of $1.5$ TeV and
a moderate CKM4 entry $V_{t^\prime b}=0.1$. By increasing the KK scale
or $V_{t^\prime b}$, the branching of $t'\to b\,W$ is enhanced, but we
observe that the decay into Higgs and $bottom$ remains well above the
20$\%$ branching in the worst case considered.
In general one can see that the flavor violating
decays  of the $t^\prime$ are significant for all parameter values
chosen, and, as long as they are kinematically allowed, they 
clearly dominate over the intuitive channel $t^\prime \to b\,W$. Of
course, the effect depends on $({R'}^2\bar{Y}^2)^2$ and will
decouple for a large enough increase of the KK scale ${R'}^{-1}$.
Therefore, which decay is dominant depends sensitively on the KK
scale $R'^{-1}$ and also on the CKM mixing $V_{t'b}$.  
In particular, for $R'^{-1}=1.5$ TeV and $V_{t^\prime b}=0.1$ (a value
favored in the fits of \cite{Alok:2010zj}), the branching ratio for
$t^\prime \to t\,h$  seems to be predicted to be dominant and about twice as large
as the one for $t^\prime \to b\, W$ over the allowed parameter space.
While for $R'^{-1}=3$ TeV and  $V_{t^\prime b}=0.1$, 
the branching ratio for $t^\prime \to t\, h$ is now 
predicted to be about two to three times smaller than that of
$t^\prime \to b\, W$. For the intermediate choice, $R'^{-1}=1.5$ TeV
and $V_{t^\prime\, b}=0.3$ the branching ratio for $t' \to b\,W$ 
 overlaps with that for  $t' \to t\,h$ over a significant range of parameter space.

In all three plots, the flavor violating decay $t'\to t Z^0$ is
subdominant, yet non-negligible either, with possible branchings ranging from
about $1\%$ to $10\%$. This channel becomes specially interesting when
the decay into Higgs is kinematically forbidden, namely for $t'$ masses
below the threshold $m_t+m_h \simeq 370$ GeV, but the decay into
$top$ and $Z$ is open.

We also include the 
suppressed decays $t' \to q_i\,h,~t' \to q_i\,Z^0$ and $t' \to
q_j\,W,~i,j=1,2$. The $Z^0$ decay width is sometimes too small and its
branching ratio falls below $10^{-3}$, which is why it does not appear
in the plot. We take a generic value for $V_{t'q_j}=0.01$ and include
FCNC coefficients  $a^u_{4i(i4)}$, $(a^u_L)_{4i(i4)}$ from our scan. 

Thus the decay  $t^\prime \to t\, h$, if kinematically allowed, is a
promising channel for observing  $t^\prime$ pair production as well as
a novel Higgs pair production channel, in the subsequent decays of the
heavy quarks. 

It may even be possible to see simultaneously the two dominant decays\footnote{
One might also be able to observe the decays $t'\to tZ^0$ even if
clearly subdominant over the parameter space.} if the branching 
ratios happen to be of similar size, giving rise to 
interesting pair production processes and decays: 

\begin{tabular}{p{7cm}p{7cm}}
\bit
\item   $pp\to t't'\to tthh$ 
\item   $pp\to t't'\to bbWW$
\eit 
&
\bit
\item   $pp\to t't'\to tbhW$
\eit\\
\end{tabular}

\noindent all potentially accessible and thus providing an indirect confirmation
(or at least a consistency check) of the warped extra dimensional
model and its parameter space. In particular, the relative importance
of these signals would provide valuable hints on the size of the KK scale as
well as of the CKM4 angle $V_{t'b}$. Note also that if the KK scale is
such that $R'^{-1}=1.5$ TeV, the lightest KK particle in the minimal scenario  would have a mass of
${\cal O}(3$ TeV) and may escape detection at the LHC, while the
exotic flavor violating decays (caused by the presence of KK particles) 
of the fourth generation quarks would still be observable.

\begin{figure}[t]
\center$
	\begin{array}{ccc}
\hspace*{-0.4cm}
	\includegraphics[width=2.2in,height=3.2in]{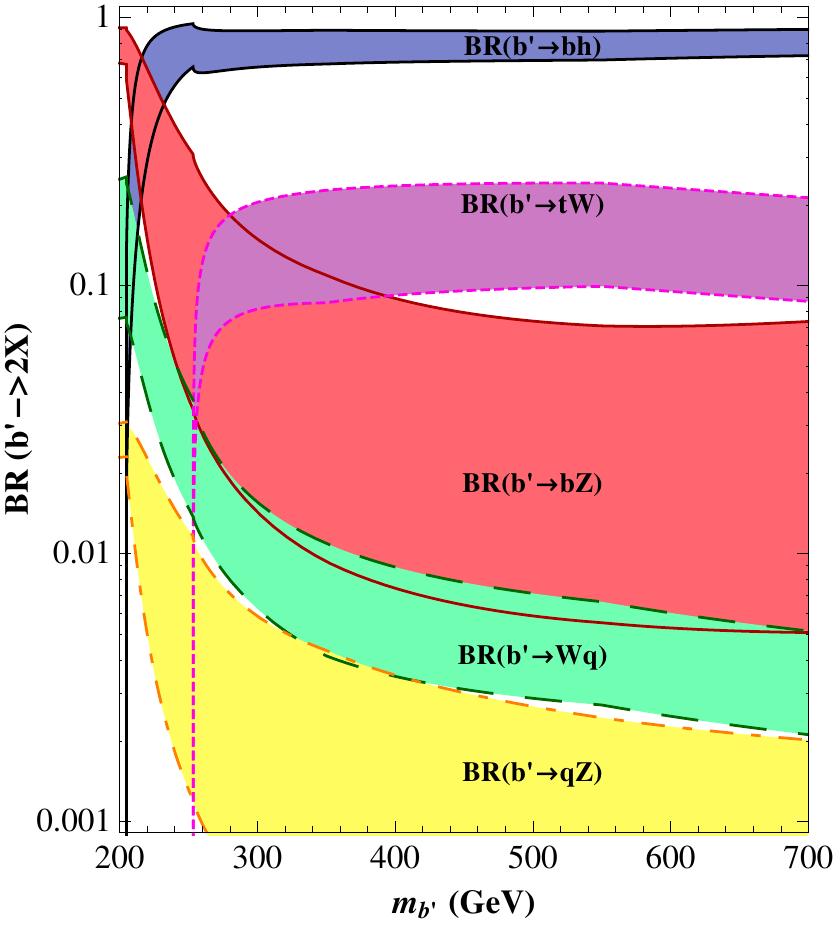}
&\hspace*{-0.2cm}
	\includegraphics[width=2.2in,height=3.2in]{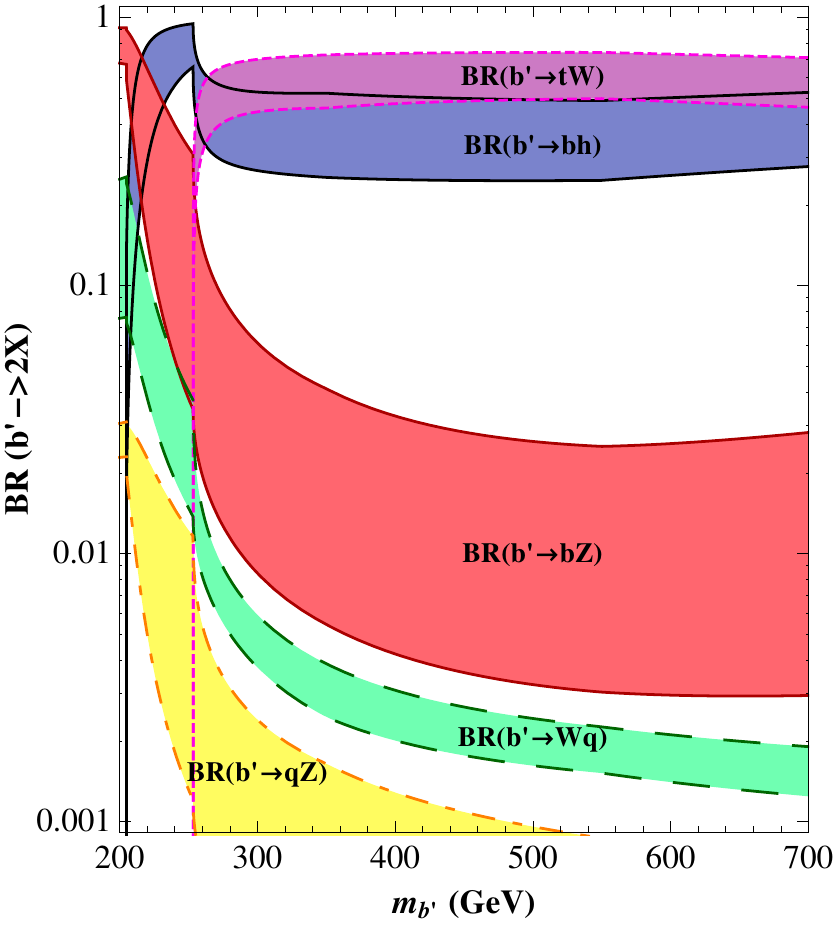}
&\hspace*{-0.2cm}
        \includegraphics[width=2.2in,height=3.2in]{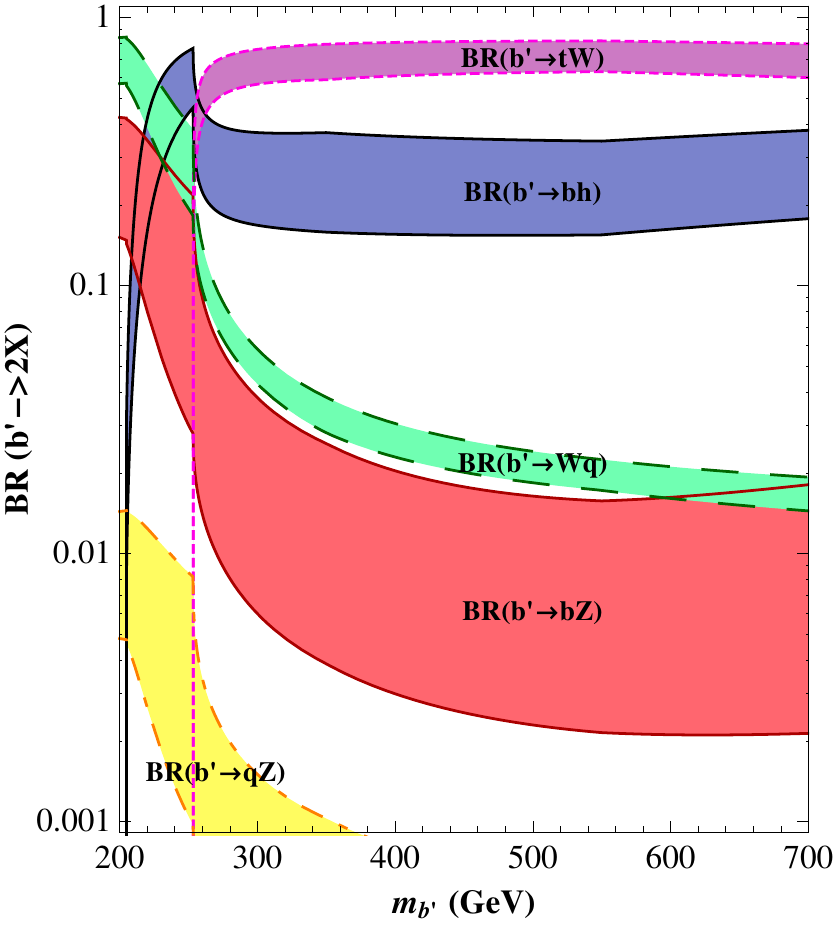}\\
        \end{array}$
\caption{Branching ratios for 2-body $b^\prime$ decays with CKM4
  mixing angle $V_{tb^\prime}=0.1$ and KK scale $R'^{-1}(\equiv M_{KK})=1.5$ TeV
  (left panel),   $V_{tb^\prime}=0.3,\ R'^{-1}=1.5$ TeV (middle panel)
  and   $V_{tb^\prime}=0.1,\ R'^{-1}=3$ TeV (right panel). We take
  $V_{cb^\prime}=V_{ub^\prime}=0.01$ throughout as well as $m_h= 200$ GeV. The bands represent 30\% likelihood for the branching ratio, according to our numerical scan, as explained in the text.}   
\label{BRbp}
\end{figure}

We perform the same analysis for the decays of the $b^\prime$ quark as
shown in Fig.~\ref{BRbp}. 
As before, we choose three parameter combinations
for the KK scale and for the main CKM4 mixing angle involved in these
decays, i.e $R'^{-1}=1.5$ TeV and $V_{t b^\prime}=0.1$, then $R'^{-1}=1.5$ TeV and
$V_{t b^\prime }=0.3$,  and finally $R'^{-1}=3$ TeV and $V_{t b^\prime}=0.1$. The
dependence of the branching ratios of FCNC decays of the  $b^\prime$
quark is more or less similar to the corresponding ones  for the 
$t^\prime$ quark, with the decay $b' \to b\,h$ dominating over all others  for $R'^{-1}=1.5$ TeV and $V_{t b^\prime}=0.1$, while for the two other parameter choices the decay $b' \to t\,W$ has the largest width for $m_{b'} \ge 250$ GeV (although it overlaps with $b' \to b\,h$ for  $V_{tb^\prime}=0.3,\ R'^{-1}=1.5$ TeV).

The flavor violating decays $b^\prime \to
b\,h$, $b^\prime \to b\,Z^0$ have a lower kinematic threshold than
$t^\prime \to t\,h$ and therefore can happen for $b'$ masses just above the Higgs
(or $Z^0$) mass. But
the W-mediated decays of the $b'$ start 
at a larger mass threshold than in the previous CKM decays of the $t'$, since
charged current decays of $b'$ will involve a $top$ quark and a $W$,
both heavy. This means that in the low $b'$ mass region, the
FCNC decays start dominating. Of course as the mixing angle
$V_{tb^\prime }$ is increased, the relative importance of the charged current
decay grows as expected. As before, we include the CKM4 and $a^d_{ij}$, $(a^d_L)_{ij}$ suppressed decays $b' \to q_i\,h,~b' \to q_i\,Z^0$ and $b' \to q_j\,W,~i,j=1,2$, with a generic value for $V_{b'q_j}=0.01$ and including the  FCNC couplings  $a^d_{4i(i4)}$, $(a^d_L)_{4i(i4)}$ from our scan.

Again, the  $b^\prime \to h\, b$ decay will be very
important in all the parameter points considered, being dominant for
low KK scale and small CKM4 mixing angles, and then competing with the
decay $b^\prime \to t\,W$ when KK scale or  $V_{tb^\prime}$ are
increased. The decay $b' \to b\,Z^0$ is suppressed relative to the
other two, but still important, with branching ratios  reaching 1\% -6\%. 

As before, we include the CKM and Yukawa suppressed decays
$b^\prime \to q'\, W$, $b'\to q h$ and $b'\to q Z$, with $q'=u,d$
and $q=s,d$. 
Again, FCNC decays of $b^\prime$ through Higgs or $Z^0$
bosons would provide an indirect indication of the warped space
scenario, even for large KK scales such as $R'^{-1}=3$ TeV.  
  From the plots one see that it may again be possible to observe at the
same time the dominant decay modes of the $b'$ quark (since these are
produced in pairs). 

For a lighter $b'$, below the threshold for $b'\to
t W$, i.e. $m_b'< 250$ GeV one sees that the FCNC decays into Higgs
and into $Z^0$ might dominate over decays into $W$ and light quarks (and hence might
substantially alter the current experimental bounds on the $b'$ mass
where CKM decays are assumed). In that situation it may be possible to
observe a mixture of events:

\begin{tabular}{p{7cm}p{7cm}}
\bit
\item   $pp\to b'b'\to bbhh$ 
\item   $pp\to b'b'\to bbZZ$
\item   $pp\to b'b'\to qWqW$
\eit 
&
\bit
\item   $pp\to b'b'\to qWbh$
\item  $pp\to b'b'\to bZbh$
\item  $pp\to b'b'\to bZqW$
\eit\\
\end{tabular}

Of these, the events containing a Higgs would be the cleanest by far,
since the Higgs itself should decay into $WW$ or $ZZ$ giving rise to
events with many gauge bosons.\\

For a heavier $b'$, it appears that two modes should dominate,
namely the FCNC decays onto Higgs and the  decays into a $W$ and a top quark
(due to our assumption of $V_{tb'}$ being the largest of the CKM4
mixing angles involved).
The possible mixed events could now be

\begin{tabular}{p{7cm}p{7cm}}
\bit
\item   $pp\to b'b'\to bbhh$ 
\item   $pp\to b'b'\to tWtW$
\eit 
&
\bit
\item   $pp\to b'b'\to tWbh$
\eit\\
\end{tabular}

All events would be easy to identify at the LHC and their relative importance 
would provide again valuable information on the model parameters of this scenario.


\bit
\item Heavy lepton decays
\eit
Once the $\tau^\prime$ lepton is produced at a collider, its FCNC
decay will proceed in the same manner as  
that of the $b^\prime$ quark. As the mass bounds on new
$\tau^\prime$ leptons and $ \nu_{\tau}'$ neutrinos are close, it may be that the
decay $\tau^\prime \to W  \nu_{\tau}'$ is kinematically forbidden, and the decay of $\tau^\prime$ to
lighter neutrinos ($\tau^\prime \to W \nu_i,\ i=1,2,3$) depends on the
specific model of  neutrino masses and mixing and may be suppressed. 
Thus the FCNC decays $\tau^\prime \to h \tau$, and
$\tau^\prime \to \tau Z^0$ could be the most dominant decays. Since we
are assuming that $m_h+m_\tau<m_{\tau'}$, the production of $\tau'$
should happen via $s$-channel W bosons and KK partners, and
therefore would typically come with associated production of
$\nu'_\tau$ (if the mixing to lighter neutrinos is smaller).

The subsequent FCNC decays of $\tau'$ should be easily disentangled at
the LHC as we would obtain several possible processes with many
leptons, such as $pp \to \tau' \nu_\tau \to \tau h W l$ for the case
of $\tau^\prime \to \tau h$ decays. The Higgs, being heavier than $200$ GeV, should
mainly decay into pairs of gauge bosons giving rise to final states of
$WWW l \tau$ or $ZZW l\tau$, i.e. three gauge bosons, one light lepton and
a $\tau$, a clean enough signal at hadron machine. These might give
rise to same-sign dilepton events, trilepton events, and pushing it,
to 6 leptons plus $\tau$ events, when every boson decays leptonically.

In the case of $\tau^\prime\to \tau Z^0$ decays, one would similarly
obtain processes like  $pp\to\tau' \nu_\tau \to \tau Z W l$. Again one
might observe same-sign dilepton events, trilepton events and when the
all bosons decay leptonically one could obtain events with four leptons and a $\tau$.

As in the previous section, a realistic analysis of these signals is
beyond the scope of this work, however, it seems clear that it would 
not be hard to disentangle them, as the branching ratios are
subdominant to $\tau' \to h \tau$, but nonetheless  significant.

\section{Conclusions and Outlook}
\label{sec:conclusion}

In this work we analyzed the effects of Higgs flavor-violating couplings in the
framework of warped extra dimensions on a fourth generation of quarks and
leptons. In this model, the Higgs Yukawa couplings are misaligned
with the fermion mass matrices, and this effects is even more pronounced in a
model with a sequential fourth fermion family, due to cumulative effects in flavor
space.

We presented both an analytical evaluation and a numerical estimate of the size of
the Higgs FCNC couplings in models with flavor anarchy. The only requirement is
that the three-generations quark masses and mixing angles should be reproduced in
the present scheme, while the fourth generations masses and mixings are allowed
to be free, limited only by $V_{CKM4}$ unitarity. We briefly discussed the
possibilities for the lepton sector, which is unfortunately complicated by the lack of
a well-defined model of neutrino masses and mixings; as well as revisited the FCNC couplings of the $Z^0$ boson with a fourth generation.

After setting up the model and evaluating the Yukawa couplings, we analyzed the
new effects on low energy FCNC observables. At tree level, the new off-diagonal
couplings affect $K^0-{\bar K}^0$, $D^0-{\bar D}^0$ and $B_{d,s}^0-{\bar
B}_{d,s}^0$ mixings. We use the data to set constraints on the $a_{ij}$, the
most stringent bound coming from $\epsilon_K$ constraining the phases of the
FCNC Yukawa couplings. The constraints are similar to the those obtained in the
three-generations scenario \cite{toharia1} and the bounds 
imposed are not stringent, even if we expect the $3 \times 3$ Yukawa couplings to be
reduced in the four-generation model. The Yukawa FCNC couplings contribute to
loop-level processes such as $b \to s \gamma$, $t \to c \gamma$, $\tau \to e,
(\mu) \gamma$ and $\mu \to e \gamma$. For the quark radiative decays, the effect
is negligible compared to SM values and $W q$  diagrams.  For leptons, depending
on the size of the FCNC Higgs Yukawa couplings, the radiative decays might
become more important and restrict the $a_{ij}^l$ beyond the expectation from
the numerical scan, especially from the $\mu \to e \gamma$ decays, and even more
as the bounds on lepton-flavor violation are expected to improve in the near
future.

As the present limits on the Higgs masses are pushed higher, especially for the
case of four generations,  the Higgs boson decay patterns can be substantially
modified  from the SM and even SM4 expectations. FCNC decay channels such as
$\tau \tau^\prime$, $b b^\prime$ and even $t t^\prime$ open for $m_h \gsim 200$
GeV, for present bounds on four-generation masses. Both $h \to \tau \tau^\prime$
 could prove to be fertile grounds for discovery of the fourth generation
leptons, if the decay $h \to \tau^\prime \tau^\prime$ is kinematically
forbidden. Similarly, the  decay $h \to b b^\prime$ could be an important
channel for $b^\prime$ discovery if off-diagonal fourth generation mixing angles $V_{u
b^\prime}, V_{c b^\prime}$ and $V_{t b^\prime}$ are small. The decays
are important for the whole parameter space $m_{t^\prime} \geq 400$
GeV, $m_{b^\prime} \geq 200$ GeV and would provide a clear indication
of the model. 

If the fourth generation quarks and leptons are heavier than the Higgs boson, their decay into
lighter quarks and Higgs bosons would be a promising channel for their discovery
and identification. In particular, the branching ratios for $t^\prime \to t h$
and $b^\prime \to b h$ compete with $t^\prime \to t Z^0$
and $b^\prime \to b Z^0$, dominate for most of the parameter space, and approach $1$ for a significant range of
$V_{t^\prime b},\ V_{tb^\prime}$ and $m_{t'},\ m_{b'}$ parameter
space. 
And the fourth generation lepton which can only decay through electroweak
processes, may not be able to decay into $W  \nu_{\tau}'$ or $W \nu_\tau$ (depending on
mass and mixing constraints in the leptonic sector), making
$\tau^\prime \to \tau h$ a dominant decay mode, and competing with $\tau^\prime \to \tau Z^0$. 

Thus, even if the KK scale is heavy, and KK particles cannot be seen at the LHC,
residual effects due to Higgs FCNC could provide the most promising indirect
signals for the warped space scenario. Our analysis shows that in a
four-generation model, which is natural in this scenario, the results could be
enhanced over the model with three generations and yield measurable signals at
the LHC.

\section{Acknowledgments}
M.T. would like to thank Kaustubh Agashe, Alex Azatov and Lijun Zhu for
many discussions regarding Higgs FCNC's in this type of scenarios. 
M.F. is grateful to Heather Logan for comments. This work was
supported in part by NSERC of Canada under SAP105354.

\section{Appendix A - Loop Calculations}
We present in this appendix explicit results for some of the radiative
corrections addressed in the main text. 
\bit
\item  $b \rightarrow s\gamma$ 
 
 The amplitude of $b \rightarrow s\gamma$ decay is
\begin{eqnarray}
\label{btosgamma}
-i\mathcal{M}=\frac{iem_{b'}\sqrt{m_{b}m_{s}}}{48\pi^{2}v_{4}^{2}}&\Big \{&(a_{24
}^{d^{*}}a_{34}^{d}P_{L}+a_{42}^{d}a_{43}^{d^{*}}P_{R})\gamma^{\mu}\Big[-2C_{00}
+ m_{b}^{2}C_{12}-m_{s}^{2}(C_{11}+C_{12}+C_{1})\nonumber\\
&+&m_{j}^{2}C_{0}+\frac{m_{s}^{2}}{m_{s}^{2}-m_{b}^{2}}(B_{0}
+B_{1})-\frac{m_{b}^{2}}{m_{s}^{2}-m_{b}^{2}}(\tilde{B}_{0}
+\tilde{B}_{1})\Big]\nonumber\\
&-&(a_{24}^{d^{*}}a_{34}^{d}P_{R}+a_{42}^{d}a_{43}^{d^{*}}P_{L})\gamma^{\mu
} m_ { s
}m_{b}\Big[C_{1}-\frac{B_{0}+B_{1}}{m_{s}^{2}-m_{b}^{2}}+\frac{\tilde{B}_{0}
+\tilde{B}_{ 1 } } { m_ { s }^{2}-m_{b}^{2}}\Big]\nonumber\\
&-&(a_{24}^{d^{*}}a_{43}^{d^{*}}P_{L}+a_{42}^{d}a_{34}^{d}P_{R})\gamma^{\mu}m_{b
}m_{b'}\Big[C_{0} -\frac{B_{0}}{m_{s}^{2}-m_{b}^{2}}
+\frac{\tilde{B}_{0}}{m_{s}^{2}-m_{b}^{2}}\Big]\nonumber\\
&+&(a_{24}^{d^{*}}a_{43}^{d^{*}}P_{R}+a_{42}^{d}a_{34}^{d}P_{L})\gamma^{\mu}m_{s
}m_{b'}\Big[C_{0}+\frac{B_{0}}{m_{s}^{2}-m_{b}^{2}}
-\frac{\tilde{B}_{0}}{m_{s}^{2}-m_{b}^{2}}\Big]
\Big \},\nonumber
\\
\end{eqnarray}
where $\tilde{B}_{0}$ and $\tilde{B}_{1}$ stand for two-point coefficient
functions with different arguments than $B_{0}$ and $B_{1}$. 
The arguments of the scalar and tensor-coefficient functions
appearing in the three-point integrals are
$\Big(P_{1}^{2},P_{2}^{2},(P_{1}+P_{2})^{2},m_{b'}^{2},m_{h}^{2},m_{b'}^{2}
\Big)$. The two-point integral coefficient functions $B_{0}$ and $B_{1}$ have the
arguments $\Big(m_{s}^{2},m_{h}^{2},m_{b'}^{2}\Big)$ while the
arguments of $\tilde{B}_{0}$ and $\tilde{B}_{1}$ are
$\Big(m_{b}^{2},m_{h}^{2},m_{b'}^{2}\Big)$. Although $\tilde{B}$'s depend on  
different arguments, their numerical values are almost the same as those of the 
$B$'s


The decay rate is
\begin{equation}
\Gamma(b\rightarrow s
\gamma)=\frac{\langle\mathcal{M}^{2}\rangle}{16\pi
m_{b}^{3}}\sqrt{m_{b}^{4}+m_{s }^{4}-2m_{b}^{2}m_{s}^{2}}.
\end{equation}

\item{$Z\to b\bar{b}$ decay}

The radiative corrections to $Z \rightarrow b\bar{b}$ vertex are 
\begin{eqnarray}
\delta g^{L}(b)=\frac{m_{b}m_{b^\prime}}{16
\pi^{2}v_{4}^{2}}&\Big \{&g_{Zb\bar{b}}^{R}
\Big[
|a_{43}^{d}|^2\Big(-2 C_{00}-m_{b}^{2}(C_{11}+C_{22}+C_{1}-C_{2})+M_{Z}^{2}
C_{12}\Big)\nonumber\\
&+&a_{34}^{d}a_{43}^{d}m_{b}m_{b^\prime}(C_{1}-C_{2})\Big]\nonumber\\
&+&g_{Zb\bar{b}}^{L}\Big[|a_{34}^{d}|^{2}\Big(-m_{b}^{2}(C_{2}+C_{1}-C_{0})+m_{
b^\prime
} ^{2}
C_{0}\Big)+a_{34}^{d\star}a_{43}^{d\star}m_{b}m_{b^\prime}(C_{2}-C_{1}
)\nonumber\\
&+&|a_{43}^{d}|^{2}\Big(-B_{1}+(m_{b}^{2}-m_{h}^{2}+m_{b^\prime}^{2})B^\prime_{0
}
\Big)+2a_{34}^{d}a_{43}^{d}m_{b}m_{b^\prime}B^\prime_{0}\Big]\Big \},
\end{eqnarray}
and
\begin{eqnarray}
\delta g^{R}(b)=\frac{m_{b}m_{b^\prime}}{16
\pi^{2}v_{4}^{2}}&\Big \{&g_{Zb\bar{b}}^{L}
\Big[
|a_{34}^{d}|^2\Big(-2 C_{00}-m_{b}^{2}(C_{11}+C_{22}+C_{1}-C_{2})+M_{Z}^{2}
C_{12}\Big)\nonumber\\
&+&a_{34}^{d \star}a_{43}^{d \star
}m_{b}m_{b^\prime}(C_{1}-C_{2})\Big]\nonumber\\
&+&g_{Zb\bar{b}}^{L}\Big[|a_{43}^{d}|^{2}\Big(-m_{b}^{2}(C_{2}+C_{1}-C_{0})+m_{
b^\prime
} ^{2}
C_{0}\Big)+a_{34}^{d}a_{43}^{d}m_{b}m_{b^\prime}(C_{2}-C_{1})\nonumber\\
&+&|a_{34}^{d}|^{2}\Big(-B_{1}+(m_{b}^{2}-m_{h}^{2}+m_{b^\prime}^{2})B^\prime_{0
}
\Big)+2a_{34}^{d \star}a_{43}^{d \star
}m_{b}m_{b^\prime}B^\prime_{0}\Big]\Big \}.
\end{eqnarray}
The results are in good agreement with \cite{Haber:1999zh} in the limit
$m_{b}\rightarrow 0$ and when the intermediate particle is the $b$ rather than
the $b^\prime$ quark. $
g_{Zb\bar{b}}^{L}$ and $ g_{Zb\bar{b}}^{R}$ are the tree level 
$Zb\bar{b}$ couplings of the SM and they are given by 
\begin{eqnarray}
 g_{Zb\bar{b}}^{L}&=&\frac{e}{\sin\theta_{W}\cos\theta_{W}}\Big(-\frac{1}{2}
+\frac{1}{3}\sin^{2}\theta_{W}\Big),\nonumber\\
g_{Zb\bar{b}}^{R}&=&\frac{e}{\sin\theta_{W}\cos\theta_{W}}\Big(\frac{1}{3}\sin^{
2}\theta_{W}\Big).
\end{eqnarray}
The arguments of the scalar and tensor-coefficient functions
\cite{Hahn:1998yk} appearing in the three-point and in the two-point integrals
are $\Big(P_{1}^{2},P_{2}^{2},(P_{1}+P_{2})^{2},m_{h}^{2},m_{b'}^{2},m_{b'}^{2}
\Big)$ and $\Big(m_{b}^{2},m_{h}^{2},m_{b'}^{2}\Big)$, respectively. We note the following:
\begin{enumerate}
 \item The largest contribution to $g^{L}(b)$ comes from the term
$\displaystyle \frac{m_{b}m_{b^\prime}}{16
\pi^{2}v_{4}^{2}}g_{Zb\bar{b}}^{L}|a_{34}^{d}|^{2}\left[m_{b^\prime}^2C_{0}
\right]
\sim 4.65202\times 10^{-6}$.
 \item The largest contribution to $g^{R}(b)$ is
$\displaystyle -\frac{m_{b}m_{b^\prime}}{16
\pi^{2}v_{4}^{2}} g_{Zb\bar{b}}^{L}|a_{34}^{d}|^{2}\left [2 C_{00} \right] \sim
-1.4136\times
10^{-5}$.
 \item In this calculation, even if some terms include phases, they contribute
as the 
coefficient of either $(C_{1}-C_{2})$, which is almost equal to zero, or
$B^\prime_{0}$ which is negligible compared to the dominant
terms. Thus, the phases in the Higgs Yukawa couplings $a_{ij}$  do not
affect the final result.  
\end{enumerate}

\item{$Z \to \tau^+\tau^-$}

The Higgs-mediated loop contribution to the width $\Gamma(Z \to l^+ l^-)$ with a
heavy $\tau^\prime$ in the loop proceeds as $Z \to b {\bar b}$ and induces a
non-universal correction to the $\tau^+ \tau^-$ decay.
However, the correction due to the FCNC in the loop is very small  $\left ( \delta g^L_{Z \tau^+\tau^-}, \ \delta g^R_{Z \tau^+\tau^-} \simeq{\cal
O}(10^{-7})\right )$ for both profiles (A) and (B) in subsection \ref{leptons})  and thus the change in width, when compared to $\displaystyle {\Gamma(\tau^+
\tau^-)}/{\Gamma(e^+ e^-)}=1.0019 \pm 0.0032$ \cite{particledata}, it does not
set any meaningful bound on the FCNC Higgs couplings in the leptonic sector.

\end{itemize}

\section{Appendix B - Fermion Masses in RS4}

First let's define our notation. If ${\bf A}$ is an $n\times n$ matrix, then
${[\bf A]}_{ij}$ represents its $\{ij\}$ first order minor, i.e. the determinant
of the $(n-1) \times (n-1)$ submatrix obtained by removing row $i$ and
column $j$ to ${\bf A}$. We will also use the notation
${\bf[A]}_{ij,\alpha\beta}$ to represent the $\{ij,
\alpha \beta\}$ second order minor of ${\bf A}$, i.e. the determinant of the 
 $(n-2) \times (n-2)$ submatrix obtained by removing rows $i$ and
$\alpha$, and columns $j$ and $\beta$ to the matrix ${\bf A}$.

\bea
M_u&=& v\ F_Q Y_u F_u,\\
&&\non\\
M_uM_u^+&=&v^2\ F_Q Y_u F^2_u Y^+_u F_Q.
\eea

We can always write 
\bea
\prod_{i=1}^{i=4} m_i = m_{t'}m_t m_c m_u= \Big|Det(M_u)\Big| =
v^4\ \Big|Det(F_Q)Det(Y_u)Det(F_u)
\Big|,
\eea
and to lowest order in ratios of $f_i$'s we can write
\bea
\!\!\!\prod_{i=2}^{i=4} m_i = m_{t'}m_t m_c = \Big|[M_u]_{{}_{11}}\Big|=v^3\
     \Big|[F_u]_{{}_{11}}[Y_u]_{{}_{11}} [F_Q]_{{}_{11}}\Big| =v^3\
f_{Q_2}f_{Q_3}f_{Q_4}
     f_{u_2}f_{u_3}f_{u_4} \Big|[Y_u]_{{}_{11}}\Big| ,\label{mcmtmtprim}
\eea
and
\bea
\!\!\!\prod_{i=3}^{i=4} m_i = m_{t'}m_t = \Big|[M_u]_{{}_{11,22}}\Big|=v^2\
     \Big|[F_u]_{{}_{11,22}} [Y_u]_{{}_{11,22}} [F_Q]_{{}_{11,22}}\Big|=v^2\
f_{Q_3}f_{Q_4}
     f_{u_3}f_{u_4} \Big|[Y_u]_{{}_{11,22}}\Big| ,
\eea
where we have used the property ${\bf [AB]=[A][B]}$.

We can therefore obtain the leading contributions to the quark masses
\bea
m_u=\frac{m_{t'} m_t m_cm_u}{m_{t'}m_tm_c}&=& v\  f_{Q_1}f_{u_1}
\frac{\Big|Det(Y_u)\Big|}{\Big|[Y_u]_{{}_{11}}\Big|}, \\
{\rm and}\hspace{3cm}
m_c=\frac{m_{t'} m_t
  m_c}{m_{t'}m_t}&=&v\
f_{Q_2}f_{u_2}\frac{\Big|[Y_u]_{{}_{11}}\Big|}{\Big|[Y_u]_{{}_{11,22}}\Big|}, 
\\ 
{\rm and} \hspace{4.6cm}
m_{t'}m_t &=& 
v^2\ f_{Q_3}f_{Q_4}f_{u_3}f_{u_4} \Big|[Y_u]_{{}_{11,22}}\Big| .
\eea

In the down sector we have

\bea
M_d&=& v\ F_Q Y_d F_d, \\
&&\non\\
M_dM_d^+&=&v^2\ F_Q Y_d F^2_d Y^+_d F_Q.
\eea

Again, we can always write 
\bea
\prod_{i=1}^{i=4} m_i = m_{b'}m_b m_s m_d= \Big|Det(M_d)\Big| =
v^4\ \Big|Det(F_Q)Det(Y_d)Det(F_d)
\Big|
\eea
and to lowest order in ratios of $f_i$'s we can write
\bea
\!\!\!\prod_{i=2}^{i=4} m_i = m_{b'}m_b m_s = \Big|[M_d]_{{}_{11}}\Big|=v^3\
     \Big|[F_d]_{{}_{11}}[Y_d]_{{}_{11}} [F_Q]_{{}_{11}}\Big| =v^3\
f_{Q_2}f_{Q_3}f_{Q_4}
     f_{d_2}f_{d_3}f_{d_4} \Big|[Y_d]_{{}_{11}}\Big|, 
\eea
and
\bea
\!\!\!\prod_{i=3}^{i=4} m_i = m_{b'}m_b = \Big|[M_d]_{{}_{11,22}}\Big|=v^2\
     \Big|[F_d]_{{}_{11,22}} [Y_d]_{{}_{11,22}} [F_Q]_{{}_{11,22}}\Big|=v^2\
f_{Q_3}f_{Q_4}
     f_{d_3}f_{d_4} \Big|[Y_d]_{{}_{11,22}}\Big| .
\eea
The leading contributions to the down quark masses are
\bea
m_d =\frac{m_{b'} m_b m_sm_d}{m_{b'}m_bm_s}&=& v\  f_{Q_1}f_{d_1}
\frac{\Big|Det(Y_d)\Big|}{\Big|[Y_d]_{{}_{11}}\Big|},  \\
{\rm and} \hspace{3cm}
m_s=\frac{m_{b'} m_b
  m_s}{m_{b'}m_b}&=&v\
f_{Q_2}f_{d_2}\frac{\Big|[Y_d]_{{}_{11}}\Big|}{\Big|[Y_d]_{{}_{11,22}}\Big|}, 
\\ 
{\rm and} \hspace{3.5cm}
m_{b'}m_b & =&
v\ f_{Q_3}f_{d_3}f_{Q_4}f_{d_4} \Big|[Y_d]_{{}_{11,22}}\Big| . 
\eea

Since $f_{Q_3}\sim f_{Q_{4}}$ we must have that
$\frac{f_{d_3}}{f_{d_4}} \sim \frac{m_{b}}{m_{b'}}\sim 10^{-2}$.

Because of this, we can find also

\bea
m_{b'}^2=v^2\ f^2_{d_4} \left( f^2_{Q_4} |Y^d_{44}|^2+ f^2_{Q_3}
|Y^d_{34}|^2\right).
\eea



\end{document}